\documentclass[graybox, secnum]{svmult}

\usepackage{mathptmx}       
\usepackage{helvet}         
\usepackage{courier}        
\usepackage{type1cm}        
\usepackage{makeidx}         
\usepackage{graphicx}        
\usepackage{multicol}        
\usepackage[bottom]{footmisc}
\usepackage{hyperref}        
\usepackage{soul}            
\hypersetup{colorlinks=true,urlcolor=blue}
\usepackage{amsmath}
\usepackage{amssymb}

\newcommand{\be}{\begin{equation}}
\newcommand{\ee}{\end{equation}}

\newcommand\p{{\bf p}}

\renewcommand\r{\rho}
\newcommand\s{{\bf s}}
\newcommand\x{{\bf x}}
\renewcommand\k{{\bf k}}
\newcommand\q{{\bf q}}

\newcommand{\e}{\eta}
\newcommand{\hMpc}{h\text{Mpc}^{-1}}

\renewcommand{\L}{\Lambda}

\newcommand{\HH}{\mathcal{H}}
\newcommand{\G}{\Gamma}

\renewcommand{\d}{\partial}
\renewcommand{\a}{\alpha}
\renewcommand{\b}{\beta}

\usepackage[square,numbers]{natbib}
\makeindex             

\begin{document}
\title*{Effective Field Theory for Large Scale Structure}
\author{Mikhail M. Ivanov 
}
\institute{Institute for Advanced Study, Einstein Drive 1, 08540 Princeton NJ, USA. \email{ivanov@ias.edu}
}
%
%
\maketitle
\abstract{~This chapter is a non-expert introduction to the 
effective field theory 
of large scale structure.
First, we give a detailed pedagogical explanation
of why previous attempts to build 
non-linear cosmological perturbation theory failed. 
After that we introduce the description of dark matter as
an effective non-ideal fluid and show how it corrects 
the shortcomings of the previous approaches.
Finally, we develop a formulation of the
effective field theory of large-scale structure 
from a nonequilibrium field 
theory perspective,
called time-sliced perturbation theory.
We show how this framework can be used for 
a consistent renormalization of cosmological correlation functions 
and a systematic resummation of large infrared effects relevant for the baryon acoustic oscillations.
}

\section*{Keywords} 
Large-scale structure, cosmological perturbation theory, effective field theory, renormalization,
baryon acoustic oscillations

\section{Introduction}

Cosmology is a rapidly evolving branch of modern physics. 
Over the last few decades cosmology has metamorphosed 
from a speculative field into a highly precise and 
predictive science.
As of now 
we have 
established a 
coherent picture of our Universe and its evolution,
with typical agreement between theory and observations at the per cent level~\cite{Planck:2018vyg}.

These spectacular advances have been driven, to a large extent, 
by the measurements of the cosmic microwave background (CMB) anisotropies.
These anisotropies are small fluctuations in temperature 
and polarization of the photons that were left over 
when first atoms were formed. The CMB anisotropies 
trace cosmological density fluctuations produced 
in a very early universe during a process called ``inflation.'' 
Inflation may have happened when typical energies 
in the Universe could be as high as $10^{16}~\text{GeV}$.
The cosmological fluctuations, thus, offer a unique way 
to probe physics at these high energies.

Importantly, the observed 
cosmological fluctuations also carry information about 
other dynamical stages of our Universe. Hence, the entire 
cosmic history can be inferred from them.
In particular, the CMB fluctuations are a sensitive 
probe of dark matter (DM), dark energy etc.

A crucial advantage of cosmological fluctuations is that 
they are small perturbations on top of a homogeneous 
and isotropic background. This means that they can be 
systematically described within linear cosmological 
perturbation theory.
That is a major reason why 
the study of the CMB has been so successful.

Despite significant progress made with the CMB, 
there are many questions that remain unanswered. 
The most pressing ones are the physical 
nature of dark matter, dark energy,
and inflation. In order to elucidate these questions 
and thus continue making progress in cosmology, 
we have
to measure more cosmological fluctuations. 
This may be problematic with the CMB as its information content 
is getting more and more exhausted.
However, there is another source of cosmological information
independent from the CMB,
which is becoming more important now. 
This is cosmic large-scale structure, 
i.e. the distributions of mass and luminous objects (e.g. galaxies),
which also trace primordial perturbations. 
At face value, large-scale structure 
offers orders of magnitude more information than the CMB
as its distribution is intrinsically three-dimensional.
In contrast, individual observables 
(Fourier modes) accessible with the CMB
are distributed across a two-dimensional celestial sphere.  

The main drawback of large-scale structure though is that 
it is a non-linear tracer of cosmological fluctuations,
and thus linear cosmological perturbation theory is 
insufficient for their study. There have been many
attempts to build a non-linear cosmological perturbation
theory for large-scale structure. But these attempts 
have not been successful because they were
based on unjustified assumptions, such 
as that dark matter behaves as a pressureless perfect fluid
in the non-linear regime. 

A major breakthrough has been made after applying particle
physics ideas to large-scale structure. 
The key observation is that the large-scale distribution
features several distance scales with a large hierarchy between them. 
In particular, the complicated  phenomena associated
with fully non-linear collapse of matter and galaxy formation
happen on scales $\sim 5~$Mpc. This is much smaller than the 
typical galaxy separations of interest~$\sim 100~$Mpc.
This means that one is working practically in a large-scale limit,
where the dynamical evolution should be 
representable in a simple way. 
In this limit it is sufficient to use only the degrees of freedom
that are
active on large scales,
whilst all effects of unknown 
short-scale physics 
can be systematically parameterized by a set of effective operators 
built out of these long-distance degrees of freedom. 
In particle physics this phenomenon is called \textit{decoupling}.
Using this key principle, one can describe the 
distribution of galaxies on large scales even without
a detailed knowledge of the galaxy formation physics.
This idea has been formalized in a framework called 
the effective field theory of large-scale structure (EFTofLSS)~\cite{Baumann:2010tm,Hertzberg:2012qn,Carrasco:2012cv} (see \cite{Baldauf:2020bsd,Cabass:2022avo} for recent reviews).
By construction, EFTofLSS is a rigorous program
of successive approximations that can be carried out to 
arbitrary accuracy. This can be contrasted with  
phenomenological models, which are constructed to merely mimic 
aspects of structure formation, but are not 
large-scale structure theories in the strict sense.

In this Chapter, we give a non-specialist introduction
to the effective field theory of large scale structure.
After giving a brief qualitative picture of structure formation
in Sec.~\ref{sec:lss}, we present a detailed pedagogical 
description of a pressureless perfect fluid model (PPFM) 
for structure formation in Sec.~\ref{sec:ppf}. 
This model is ultimately wrong, 
but it will give us many valuable lessons.
In particular, we will see how IR and UV singularities
in PPFM are related to physical effects 
of large-scale bulk flows and short-scale
backreaction. We introduce the EFTofLSS in terms of Eulerian
hydrodynamics in Sec.~\ref{sec:eftlss}, 
where we show in detail how this approach
corrects the shortcomings of PPFM.  
In Sec.~\ref{sec:tspt} we present 
a path integral formulation for the EFTofLSS,
which proved particularly convenient 
for the study of baryon acoustic oscillations.
In Sec.~\ref{sec:data} we compare the 
predictions of the EFTofLSS for the dark matter power 
spectrum with 
N-body simulation data and find perfect agreement 
on mildly-nonlinear scales. We briefly discuss
extensions to galaxy bias and redshift space distortions 
in Sec.~\ref{sec:rsd}. We draw conclusions in Sec.~\ref{sec:sum}.

\section{Why does the EFT of LSS work?}\label{sec:lss}

Large scale structure of the universe is the distribution of matter and 
galaxies on large cosmological scales. This distribution has been 
observed for a very long time, much before the CMB
The galaxies are clearly not distributed randomly. 
In fact, the structure of our Universe on large scales
is a result of cosmological evolution that depends on the 
initial conditions and constituents of our Universe. 
This distribution is approximately homogeneous and isotropic 
on very large scales, which suggests that its proper 
description should be in terms of small 
fluctuations around a smooth cosmological background. 
This matches observable properties of other 
cosmological fields, such as the CMB temperature fluctuations. 

The CMB temperature fluctuations are very small, $\delta T/T\sim 10^{-5}$,
and hence they can be accurately described by means of 
relativistic linear 
cosmological perturbation theory. 
Just like the CMB, galaxies represent fluctuations 
in the cosmological density field.
However, unlike the CMB, they cannot be described
with linear theory.
Galaxies are produced by 
complicated astrophysical 
processes taking place on the background of 
collapsing matter. This collapse produces halos
where galaxies reside.
As a result, galaxies probe the underlying  
matter distribution indirectly,
i.e. galaxies are \textit{tracers} of matter. 
The clustering of matter is also nonlinear 
as it subject to Jeans instability.

\begin{figure}
\centering
\includegraphics[width=0.69\textwidth]{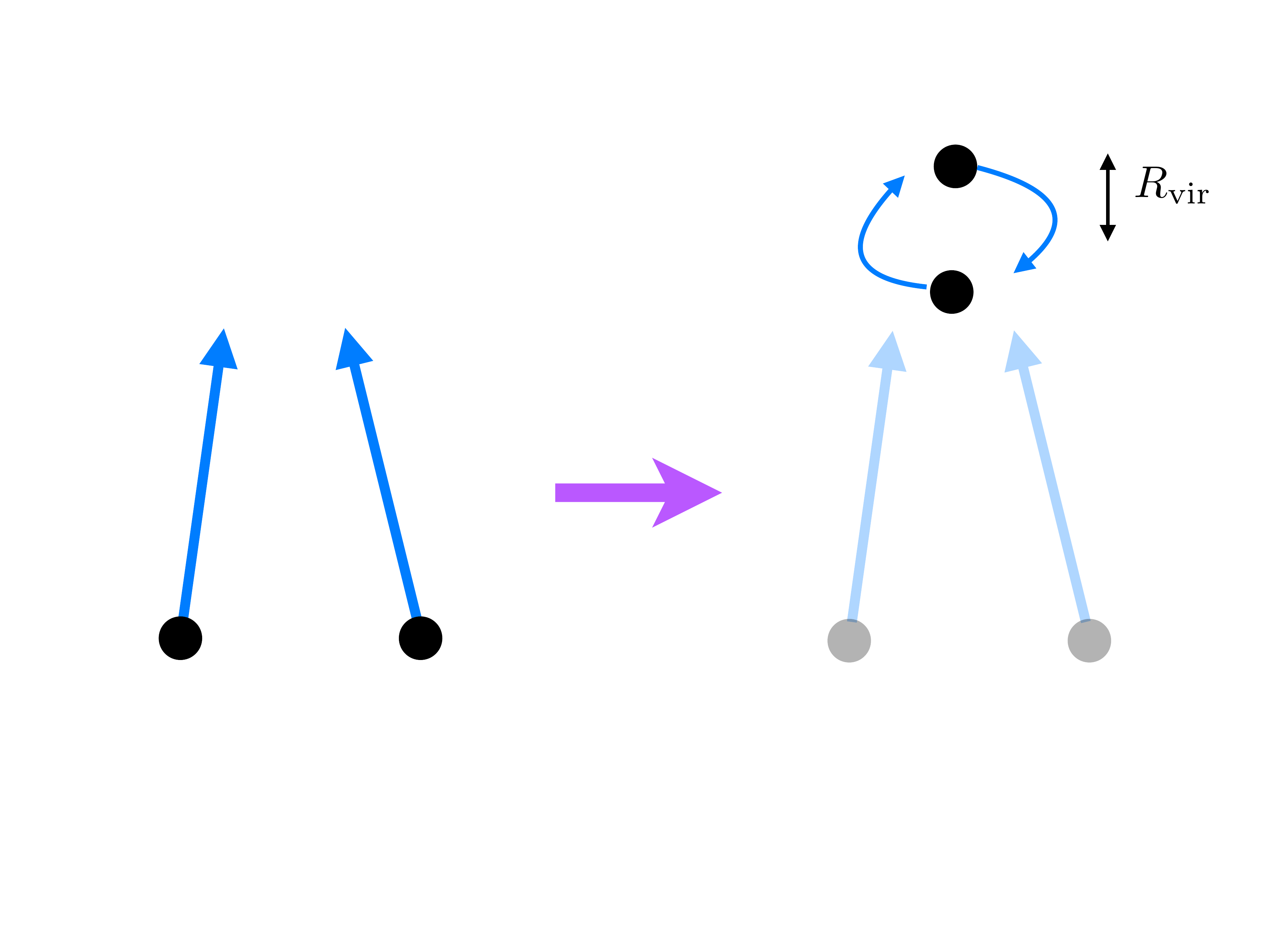}
\caption{
Cartoon of the DM clustering. \textit{Left panel:}
collapse of DM particles. When the particles approach 
each other sufficiently close, they ``stick'' together by 
gravity and form a virialized object, DM halo (\textit{Right panel}).
$R_{\rm vir}$ is the typical virial radius of the halo.
}
\label{fig:cartoon}
\end{figure}

In order to obtain a simple physical picture 
of matter clustering, we can think of matter 
as a collection of collisionless particles. 
These particles move with certain velocities set by
the initial conditions in our Universe, 
and if they get sufficiently close to one another, 
they virialize and form bound structures -- halos, 
see Fig.~\ref{fig:cartoon}.
In this simplified picture DM particles 
move in a mean gravitational potential 
until they stick together by
gravity forces at short scales.

There are two relevant scales in this problem:
the Hubble horizon scale $l_H\sim \mathcal{H}^{-1}\sim 3$ $h^{-1}$Gpc and 
the halo virial radius $R_{\rm vir}\sim 3$ $h^{-1}$Mpc. 
These scales correspond to different physics: 
at the horizon the relativistic effects become important
and one has to use general relativity to describe motions 
on these scales. 
At the scales of the halo radius
the fully non-linear processes of gravitational collapse take place,
which are hard (if possible) to model analytically. 
However, there is a wide range of scales $r$,
\be  
\label{eq:scales}
R_{\rm vir} \ll r \ll l_H\,,
\ee
where one can build a systematic analytic description of the nonlinear 
galaxy distribution 
in terms of the distance (derivative) expansion. 
These scales are called quasi-linear, 
because their description is formulated in terms
of a perturbation theory over the linear theory result.
On one end, physics operating on these scales is essentially Newtonian,
since the relativistic corrections are suppressed by powers of $r/l_H$.
On the other end, the fully nonlinear effects of halo formation
and virialization
are suppressed by powers of $R_{\rm vir}/r$.
In the EFT jargon, the halo formation physics ``decouples''
from the large-scale physics.
The typical galaxy separations that are relevant for galaxy surveys 
are $r\sim (10 - 100)~h^{-1}$Mpc, so that the relativistic effects 
can be ignored.
However, the non-linear effects associated with the 
expansion in $R_{\rm vir}/r$ are very important
in practice.
The EFTofLSS
is a systematic framework that allows to 
compute the galaxy distribution perturbatively,
effectively 
expanding in powers of the small parameter $R_{\rm vir}/r$. 

In this Chapter, we give an introduction into this subject,
focusing on the physical intuition behind this approach 
and the most important technical aspects.
In what follows, we assume that the reader possess basic knowledge
of 
cosmological perturbation theory 
and quantum field theory.

\section{A Wrong but Useful Model: Pressureless Perfect Fluid}\label{sec:ppf}

There are three key empirical facts about dark matter: 

\begin{enumerate}

\item DM behaves like dust on large scales, i.e. its pressure
is vanishingly small

\item DM perturbations are small on large scales

\item The initial conditions for structure formation 
in our Universe are Gaussian to a very good accuracy

\end{enumerate}

These three facts are consistent with  
linear cosmological perturbation theory,
where DM fluctuations 
are described as a pressure-less perfect fluid. 
Therefore, it seems reasonable to extend 
this description to the fully non-linear level. 
Historically, this had been the traditional approach
before the EFTofLSS.
The pressureless perfect fluid approach is wrong, but however, 
it gives us many useful lessons and important technical tools. 
Thus, in this section we will consider the 
pressureless perfect fluid
hydrodynamics as a toy model 
for matter clustering.
This approach is also known as standard perturbation theory~\cite{Bernardeau:2001qr}.
Our main variables will be  the peculiar velocity field $v^i$
and the overdensity field 
\be
\delta (\tau,\x)=\frac{\rho(\tau,\x) -  \rho_0(\tau)}{\rho_0(\tau)} \,, 
\ee
where $\rho_0$ is the background density,
and $\tau$ is conformal time. The naive 
perturbation theory assumption is that $\delta$
and $|\vec{v}|$ are small parameters.

The equations of motion for dust on the 
cosmological background in the Newtonian approximation
(modes are sufficiently inside the horizon) are given by 
\be 
\label{eq:hydro0}
\begin{split}
& \frac{\d}{\d \tau} \delta  
+
\d_i[(1+\delta )v^i ] = 0\,,\\
& \frac{\d}{\d \tau}v^i + \mathcal{H} v^i 
+
v^j  \d_j v^i 
 = -
 \d_i \Phi  \,,
\end{split}
\ee
where the perturbed gravitational potential satisfies the 
Poisson equation
\be
\Delta \Phi  = 4\pi G \rho_0 a^2 \delta =\frac{3}{2}\Omega_m \HH^2 \delta \,,
\ee
and $\Omega_m(\tau)$ is the time-dependent 
dark matter density fraction, 
\[
\Omega_m(\tau)=\Omega_m^{(0)}a^{-3}(\tau)/(\Omega_m^{(0)}a^{-3}(\tau)+\Omega^{(0)}_\Lambda)\,,
\]
while $\Omega_m^{(0)},\Omega^{(0)}_\Lambda$ are density
fractions at the current conformal time $\tau_0$.

The only sources of non-linearity in Eqs.~\ref{eq:hydro0}
are convective derivatives e. g. $ v^j  \d_j v^i $.
If we drop these terms for the time being, 
we will find the standard linear equations of motion 
that describe the 
cosmological growth of structure. Let us discuss this 
linear solution in detail.

\subsection{Linear Theory}

Looking at linearized Eq.~\eqref{eq:hydro0}
we notice that the source term in the r.h.s. is a gradient 
of the gravitational potential, 
which can only source the longitudinal 
part of $v^i$. Thus, we can neglect the curl part
in what follows. 
Introducing the velocity 
divergence field
\be 
\theta = -\frac{\d_i v^i}{\HH}\,,
\ee
we can rewrite the Euler and continuity equations as 
\be 
\label{eq:hydro_lin2}
\begin{split}
& \HH^{-1} \d_\tau \delta  -  \theta  = 0\,,\\
& \HH^{-1} \d_\tau  \theta
+ \left(1+ \frac{\d_\tau \HH}{\HH^2} \right)\theta
-\frac{3}{2} \Omega_m(\tau) \delta  
 = 0 \,.
\end{split}
\ee
The solution to this equation depends on the background evolution. 
During matter domination, $\Omega_m(\tau)=1$, $\HH=\frac{2}{\tau}$,
the cosmologically relevant solution is given by the growing mode,
\be
\label{eq:gmode0}
\theta  = \delta = a(\tau )\delta_0(\x)= \delta^{(1)}(\tau,\x)  \,,
\ee
where $\delta_0(\x)$ is the density initial 
fluctuation. In a more general situation we have
\be
\label{eq:gmode}
 \delta^{(1)}(\tau,\x)  = D_+(\tau ) \delta_0(\x)\,,\quad \text{where}
 \quad D_+(\tau)=\frac{5}{2}\Omega_m^{(0)}\HH_0^2 \frac{\HH}{a}\int_0^\tau d\tau'\frac{a(\tau')}{\HH'^2(\tau')}\,,
\ee
where $\HH_0^2 = \HH^2(\tau_0)$.

\begin{figure}
\centering
\includegraphics[width=0.49\textwidth]{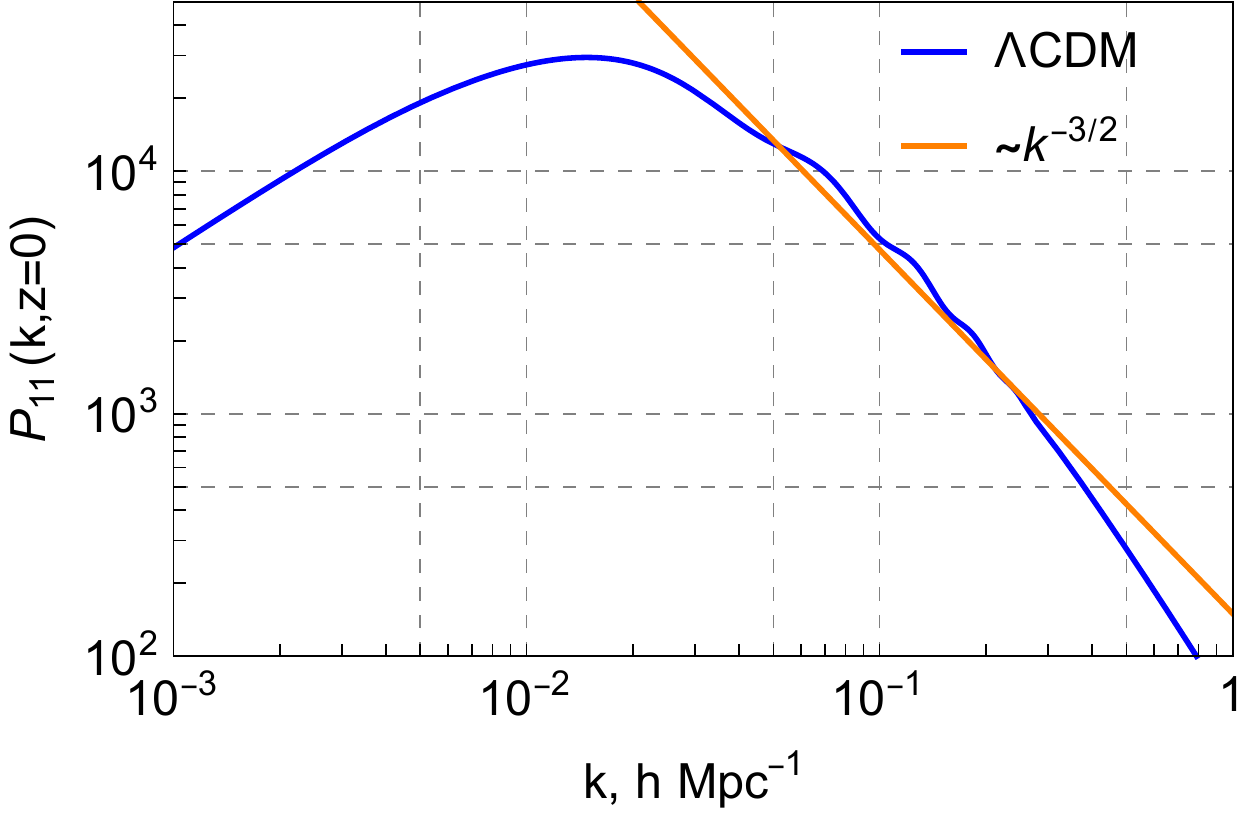}
\includegraphics[width=0.49\textwidth]{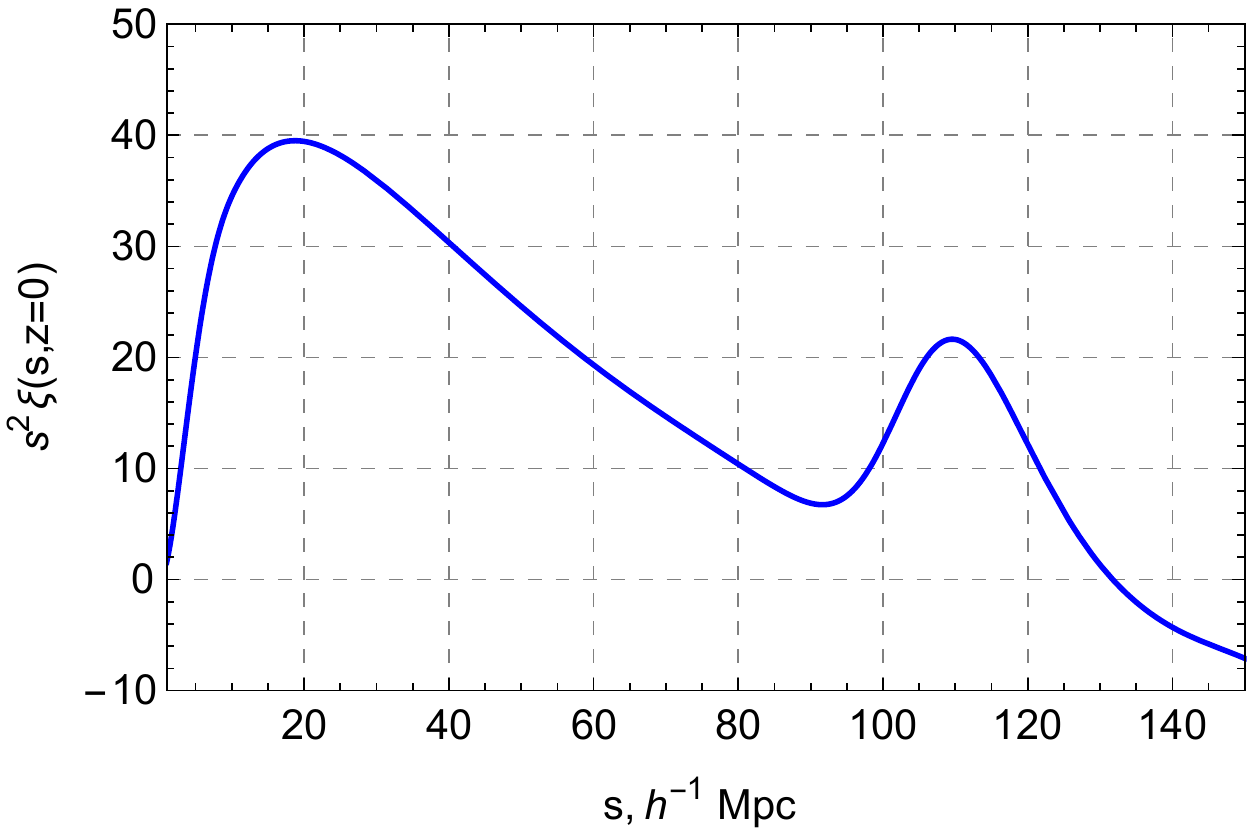}
\caption{
Left panel: the linear matter power spectrum $P_{11}$ at redshift $z=0$
in the $\L$CDM model (blue curve) vs. the power-law approximation 
for $k\approx 0.1~h~\text{Mpc}^{-1}$. 
Right panel: linear two-point correlation function.
}
\label{fig:Pk}
\end{figure}

We will work in Fourier space\footnote{Our convention 
is $\delta(\x)=\int_\k f_\k e^{i\k\cdot \x}$, $\int_\k\equiv \int\frac{d^3k}{(2\pi)^3}$, and $\delta(\k)=\int d^3x~\delta(\x)e^{-i\k\cdot \x}$.} where
the initial conditions are defined by the field 
$\delta_0(\k)$. This field is a stochastic 
variable. 
This reflects an important feature that
the 
matter distribution is stochastic, i.e. the observed galaxy maps (tracing matter)
can be thought of as drawn from a certain distribution.
It is the statistical properties of this 
distribution that carry cosmological information
and can be compared to model predictions. 
The simplest such 
statistic is the two-point function of the 
matter density, 
\be
\langle \delta_0(\k) \delta_0(\k')\rangle = (2\pi)^3 \delta^{(3)}_D(\k+\k') P_{\rm lin}(k)\,.  
\ee
Hence, in linear theory we have 
\be
\langle \delta^{(1)}(\k) \delta^{(1)}(\k')\rangle = 
(2\pi)^3 \delta^{(3)}_D(\k+\k')P_{11}(k,\tau)\,,
\ee
where $P_{11}(k,\tau)= D_+^2(\tau)P_{\rm lin}(k)$. As standard 
in cosmological perturbation theory, we choose the normalization such that $D_+(\tau=\tau_0)=1$,
in which case $ P_{\rm lin}(k)$
is a liner matter power spectrum at redshift zero.
We show it as a function of wavenumber in Fig.~\ref{fig:Pk}
for the WMAP-like cosmology used in the simulations~\cite{Kim:2011ab}.

The linear matter power spectrum in the $\Lambda$CDM model has an intricate 
shape that reflects different stages in the cosmological 
evolution: it has asymptotics
$P_{11}\propto k~(k\to 0)$, $P_{11}\propto k^{-3}\ln^2(k)~(k\to \infty)$,
and a peak at the wavenumber that corresponds to the 
comoving size of
Hubble horizon at the time of the matter-radiation equality, $k_{\rm eq}\approx 0.015~h~\text{Mpc}^{-1}$. 
In a narrow range of scales the 
matter power spectrum can be approximated as a power-law~\cite{Carrasco:2013mua,Pajer:2013jj},
\be
\label{eq:power}
 P_{\rm lin}(k) \approx \frac{1}{2\pi^2 k_{\rm NL}^3}\left(\frac{k}{k_{\rm NL}}\right)^{n}\,,
\ee
see Fig.~\ref{fig:Pk}.
From most observationally relevant wavenumbers are $k\sim 0.1\hMpc$,
where $n\approx -1.5$ provides a good approximation to the actual 
power spectrum.
We stress that we will use this approximation
only for rough estimates. All actual 
calculations will be based on the actual $\Lambda$CDM power spectrum.

Another important quantity is the position space correlation 
function, which is a Fourier image of the power spectrum,
\be
\langle \delta(\x)\delta(\x+\s)\rangle \equiv \xi(s) = \frac{1}{2\pi^2}\int dk~k^2
P_{\rm lin}(k)\frac{\sin(ks)}{ks}\,.
\ee
The correlation function for the $\Lambda$CDM cosmology 
is shown in Fig.~\ref{fig:Pk}. Note the bump at $s_{\rm BAO}\sim 110~h^{-1}$Mpc,
which corresponds to the baryon acoustic oscillations in the matter power spectrum. We see that 
the correlation function is a smooth function of scales, and it can be approximated
as a power law in position space plus a Gaussian that roughly captures 
the BAO peak.

Let us estimate now what is the typical 
density variance at a given point $\x$.
Quite surprisingly, we find that it is actually infinite
for the $\Lambda$CDM power spectrum,
\be  
\langle\delta^2(\x)\rangle=\int \frac{d^3q}{(2\pi)^3}P_{\rm lin}(q) \to \infty \,.
\ee
This integral diverges in the UV, and 
hence, the assumption that $\delta$
is small is broken down by short scale mass fluctuations
that are formally infinite in the $\Lambda$CDM cosmology. 
It seems that our linear theory 
is not actually well defined at all.

A more reasonable 
approach would be to remove the short-scale 
fluctuations first, i.e. 
introduce a low pass filter, e.g. a spherical top-hat
in position space, 
\be
W_R(\x) = \frac{3}{4\pi}H(|\x|-R)\,, 
\ee
which effective removes all fluctuations 
from scales that are shorter than $R^{-1}$. 
Note that $H(x)$ above is the Heaviside step function.
The filtered density field is given by 
$\delta^{(1)}(\k)\tilde{W}_R(kR)$, where $\tilde{W}_R(kR)={3j_1(kR)}/{(kR)}$
is the Fourier transform of the top hat.
The mass variance
inside a sphere of the radius $R$ 
is given by
\be  
\langle\delta^2_R(\x)\rangle=\int \frac{d^3k}{(2\pi)^3}P_{\rm lin}(k) |\tilde{W}_R(kR)|^2\,.
\ee
By appropriately choosing the filtering scale 
$R$ we can make sure that the coarse-grained
density fluctuations are under control.
This observation suggests that 
natural variables in our perturbation theory
should actually be \textit{smoothed} density 
and velocity fields. Hence, we need, in fact, to smooth 
equations \eqref{eq:hydro0}, which is not done 
in the PPMF. 
This is the first observation that
signals the inconsistency of our toy model.

Another quantity of interest is the displacement field,
which measures how much does a fluid
element positioned at $\x$
traveled over the structure formation 
history. Defining  $\x=\x_{ini}+\textbf{s}(\x_{ini})$,
we find 
\be
\s =  \int_{\tau_{ini}}^{\tau_0}  d\tau ~\textbf{v}(\tau,\x)\,.
\ee
The displacement field is also 
stochastic since it is sourced by 
$\textbf{v}$, which is a random stochastic variable.
Using the linear theory result 
$\d_i v^i = - a \mathcal{H} \delta_0 $
, we find that variance of the displacement field is given by 
\be\label{eq:dispvar}
\langle s_i(\x) s^i(\x) \rangle =  \int_\q \frac{P_{11}(q) }{q^2}\,.
\ee
Unlike the mass fluctuation variance, the 
displacement variance in our universe is finite because the above
integral actually converges for the 
$\Lambda$CDM power spectrum, 
\be 
\label{eq:displvar}
\langle s_i(0) s^i(0) \rangle \approx  [6~h^{-1}\text{Mpc}]^2\,.
\ee
However, if the power spectrum were a power low 
$P_{11}\propto k^{n}$, it would be divergent 
in the IR for $n\leq -1$ and in the UV for $n> -1$.
Hence, the fact that the all-scale displacement variance is finite 
in our universe is simply a coincidence.

\subsection{Non-linear perturbation theory}

Let us now solve Eqs.~\eqref{eq:hydro0} peturbatively. 
As before, we will replace the velocity
field with its divergence.
Let us first recast Eqs.~\eqref{eq:hydro0_f}
as second order differential equations on $\delta$
and $\theta$ and in Fourier space,
\be 
\begin{split}
& \HH^2 \left[-a^2 \d_a^2 +\frac{3}{2}(\Omega_m-2)a\d_a
+\frac{3}{2}\Omega_m\right]\delta_\k = \HH^2\mathcal{S}_\beta 
+\HH \d_a(a\HH \mathcal{S}_\alpha)\,,\\
& \HH^2
\left[a^2 \d_a^2 
+\left(4-\frac{3}{2}\Omega_m\right)a\d_a+ 
\Bigg(2
+\frac{\d_a^2\HH}{\HH}  +
\left(4-\frac{3}{2}\Omega_m\right)\frac{\d_a\HH}{\HH}
-3\Omega_m\Bigg) \right] \theta_\k \\
&\quad \quad \quad \quad = 
-\d_a(a \HH^2 \mathcal{S}_\beta)-\frac{3}{2}\Omega_m\HH^2\mathcal{S}_\alpha \,,
\end{split}
\ee
where the sources are given by 
\be 
\begin{split}
 & \mathcal{S}_\alpha =  -\int_{\q_1\q_2}\delta^{(3)}_D(\k-\q_{12})
 \a(\q_1,\q_2) \theta_{\q_1}\delta_{\q_2}\,,\\ 
 & \mathcal{S}_\beta = -\int_{\q_1\q_2}
\delta^{(3)}_D(\k-\q_{12})
\b(\q_1,\q_2)  \theta_{\q_1} \theta_{\q_2}\,,
\end{split}
\ee
and we introduced the nonlinear
kernels
\be
\a(\q_1,\q_2)\equiv  \frac{\q_{12}\cdot \q_1}{q_1^2}\,,\quad \beta(\q_1,\q_2)
=\frac{\q^2_{12} (\q_1\cdot \q_2)}{2q_1^2q_2^2}\,.
\ee
The retarded Green's function of $\delta$
and $ \theta$,
which can be obtained by solving the above equitations with 
the source terms replaced by $\delta(a-a')$, is given by~\cite{Baldauf:2014qfa,Baldauf:2020bsd}
\be
\label{eq:green}
G_\delta(a,a') =H(a-a') \frac{2}{5}\frac{1}{\HH_0^2\Omega_m^{(0)}}
\frac{D_+(a')}{a'}\left(\frac{D_-(a)}{D_-(a')} -\frac{D_+(a)}{D_+(a')}\right)\,,
\ee
and $G_{\theta} = -\frac{d\ln D_+}{d\ln a}(a)
G_\delta$, 
where $D_-=\HH/(a\HH_0)$ is the decaying mode and $H(x)$ is
the Heaviside step function.

It is convenient now to switch to 
a new time variable $\eta \equiv \ln D_+(\tau)$,
and the rescaled velocity divergence field 
\be
\Theta = \frac{\theta}{f} = -\frac{\d_i v^i}{f\HH}\,,\quad \text{where}
\quad f\equiv \frac{d\ln D_+}{d\ln a}\,.
\ee
This choice allows us to minimize time-dependence in the equations~\cite{Blas:2015qsi}. 
After these transformations Eqs.~\eqref{eq:hydro0}
take the following form in Fourier space:
\be 
\label{eq:hydro0_f}
\begin{split}
 & \d_\eta \delta_\k -\Theta_\k = 
\int_{\q_1\q_2}\delta^{(3)}_D(\k-\q_{12})
 \a(\q_1,\q_2) \Theta_{\q_1}\delta_{\q_2}\,,\\
& \d_\eta \Theta_\k -\frac{3\Omega_m}{2f^2}\delta_\k+\left(
\frac{3\Omega_m}{2f^2}-1
\right)\Theta_\k = \int_{\q_1\q_2}
\delta^{(3)}_D(\k-\q_{12})
\b(\q_1,\q_2)  \Theta_{\q_1} \Theta_{\q_2}\,.
\end{split}
\ee
In a matter dominated universe $\Omega_m=f=1$, so the coefficients
in the l.h.s. of the above equations are time-independent.
In the actual $\Lambda$CDM cosmology it turns out that $\Omega_m\approx f^2$
with a very good accuracy for all times. Therefore, it is customary to 
develop a perturbative scheme in which  
$\Omega_m/ f^2=1$ at zeroth order. This approach is known as the ``Einstein-de-Sitter'' (EdS) approximation~\cite{Blas:2013bpa,Blas:2015qsi,Chudaykin:2020aoj}. 
In what follows we will
use this approximation  and
stick to the zeroth EdS order.

To solve Eq.~\eqref{eq:hydro0_f} perturbatively, we
introduce the following series ansatz, 
\be
\label{eq:ans_spt0}
\begin{split}
\delta(\k,\tau)= D^n_+(\eta)\sum_{n=1}^{\infty}\delta^{(n)}(\k)\,,\quad 
\Theta(\k,\tau)= D^n_+(\eta)\sum_{n=1}^{\infty}\Theta^{(n)}(\k)\,,
\end{split}
\ee 
where the $n$'th order perturbative solution is given by
\be
\label{eq:ans_spt}
\begin{split}
\delta^{(n)}(\k) = \left[\prod_{i=1}^n \int_{\q_i}\right]\delta_D^{(3)}(\k-\q_{1...n})F_n(\q_1,...,\q_n) 
\left[\prod_{i=1}^n\delta_0(\q_n)\right]\,,\\
\Theta^{(n)}(\k) = \left[\prod_{i=1}^n \int_{\q_i}\right]\delta_D^{(3)}(\k-\q_{1...n})G_n(\q_1,...,\q_n) 
\left[\prod_{i=1}^n\delta_0(\q_n)\right]\,,
\end{split}
\ee 
and $F_n,~G_n$ are some yet unspecified momentum-dependent kernels. 
Demanding that this ansatz solve
Eqs.~\eqref{eq:hydro0_f} at any given order in $\delta_0$,
these equations then transform into
recursion relations for the kernels $F_n, G_n$,
\be 
\begin{split}
F_n(\q_1,...\q_n) = &  \sum_{m-1}^{n-1}\frac{G_m(\q_1,...,\q_m)}{(2n+3)(n-1)}
[(2n+1)\alpha(\q_{1m},\q_{(m+1)...n})F_{n-m}(\q_{m+1},...\q_n)\\
&+2\beta(\q_{1m},\q_{(m+1)n})G_{n-m}(\q_{m+1},...\q_n)]\,,\\
G_n(\q_1,...\q_n) = &  \sum_{m-1}^{n-1}\frac{G_m(\q_1,...,\q_m)}{(2n+3)(n-1)}
[3\alpha(\q_{1m},\q_{(m+1)...n})F_{n-m}(\q_{m+1},...,\q_n)\\
&+2n\beta(\q_{1m},\q_{(m+1)...n})G_{n-m}(\q_{m+1},...,\q_n)]\,.
\end{split}
\ee
In particular, we find $F_1=G_1=1$ (recovering the linear growing mode), and 
\be
\begin{split}
F_2(\q_1,\q_2) = \frac{5}{7}+\frac{(\q_1\cdot \q_2)}{q_1q_2}\left(
\frac{q_1}{q_2}+\frac{q_2}{q_1}
\right) +\frac{2}{7}\frac{(\q_1\cdot\q_2)^2}{q_1^2 q_2^2}\,,\\
G_2(\q_1,\q_2)  = \frac{3}{7}+\frac{(\q_1\cdot \q_2)}{q_1q_2}\left(
\frac{q_1}{q_2}+\frac{q_2}{q_1}
\right) +\frac{4}{7}\frac{(\q_1\cdot\q_2)^2}{q_1^2 q_2^2}\,.
\end{split}
\ee 

Note that the kernels $F_n$ and $G_n$ have IR singularities 
when some of its arguments are soft. If we keep $l$ momenta fixed
and send the remaining $n-l$ uniformly to zero, we will get 
\be
\label{eq:poles}
\lim_{\varepsilon\to 0} F_n(\k_1,...,\k_l,\varepsilon\q'_{l+1},...,\varepsilon \q'_{n})=\mathcal{O}(\varepsilon^{l-n})\,,
\ee
and the same for $G_n$. We will return to these IR singularities
later.

Using Eq.~\eqref{eq:ans_spt}, we can compute now various 
correlation functions in perturbation theory. 
The power spectrum at order $[\delta_0]^4$ is given by 
\be
\begin{split}
& \langle \delta_\k \delta_{\k'} \rangle =\langle (\delta^{(1)}_\k+\delta^{(2)}_\k+\delta^{(3)}_\k) (\delta^{(1)}_{\k'}+\delta^{(2)}_{\k'}+\delta^{(3)}_{\k'}) \rangle \\
& = 
\langle \delta^{(1)}_\k \delta^{(1)}_{\k'} \rangle
+ \langle \delta^{(2)}_\k \delta^{(2)}_{\k'} \rangle
+2\langle \delta^{(3)}_\k \delta^{(1)}_{\k'} \rangle
 \\
 &=  
(2\pi)^3 \delta_D^{(3)}(\k+\k')( P_{11} + 
P_{13}+P_{22})\,,
\end{split}
\ee
where we used the exchange symmetry $\k\longleftrightarrow \k'$
of the $P_{13}$ correction.
In what follows we will use primes to denote 
stripping off Dirac delta-functions from statistical averages, e.g.
\be 
\langle \delta_\k \delta_{\k'} \rangle' = P(k)\,.
\ee
The combinations like $\langle \delta^{(2)}_\k \delta^{(1)}_{\k'} \rangle$ vanish for Gaussian initial conditions because 
they contain an odd number of the initial density fields,
e.g. $\langle \delta_0(\k_1) \delta_0(\k_2) \delta_0(\k_3) \rangle = 0$.
The statistical averages over $\delta_0$
generate the following convolution integrals:
\be
\begin{split}
& P_{22}(k)=\langle \delta^{(2)}_\k \delta^{(2)}_{\k'}\rangle'=2 \int_\q [F_2(\q-\k,\q)]^2 P_{11}(q)P_{11}(|\k-\q|)\,,\\
&P_{13}(k)=2\langle \delta^{(1)}_\k \delta^{(2)}_{\k'}\rangle'=6 P_{11}(k) \int_\q F_3(\k,\q\,-\q) P_{11}(q)\,.
\end{split}
\ee
These integrals have some physical interpretation. 
When $q \gg k$,
$P_{22}$ describes how
two short-scale modes with wavevectors 
$\q$ and $\k-\q$ produce a long-wavelength
mode with wavevector $\k$.
In contrast, the correction $P_{13}$ captures the modulation 
of the existing mode $\k$ by a mode $\q$. 
We will discuss the physical effect 
of these corrections shortly when we consider 
IR and UV limits of these loop integrals.

Another important aspect of non-linear structure formation
is non-Gaussianity~\cite{Scoccimarro:1997st,Scoccimarro:2000sn,
Baldauf:2014qfa,Eggemeier:2018qae,Ivanov:2021kcd,Philcox:2022frc}. Indeed, we see that in perturbation theory 
the distribution of the overdensity field acquires 
higher order correlations. The simplest one is the 3-point correlation,
captured by the bispectrum, 
\be 
\langle \delta(\k_1)
\delta(\k_2)
\delta(\k_3)
\rangle \equiv (2\pi)^3 \delta_D^{(3)}(\k_{123})B(\k_1,\k_2,\k_3)\,.
\ee 
Using the perturbative expansion~\eqref{eq:ans_spt0} we obtain, at 
order $[\delta_0^4]$:
\be
\begin{split}
B(\k_1,\k_2,\k_3) = & \langle 
\delta^{(2)}_{\k_1}
\delta^{(1)}_{\k_2} 
\delta^{(1)}_{\k_3}
\rangle' +\text{2 cyc.} = 2F_2(\k_1,\k_2)P_{11}(k_1)P_{11}(k_2) +\text{2 cyc.}
\end{split}
\ee

\begin{figure}
\centering
\includegraphics[width=0.99\textwidth]{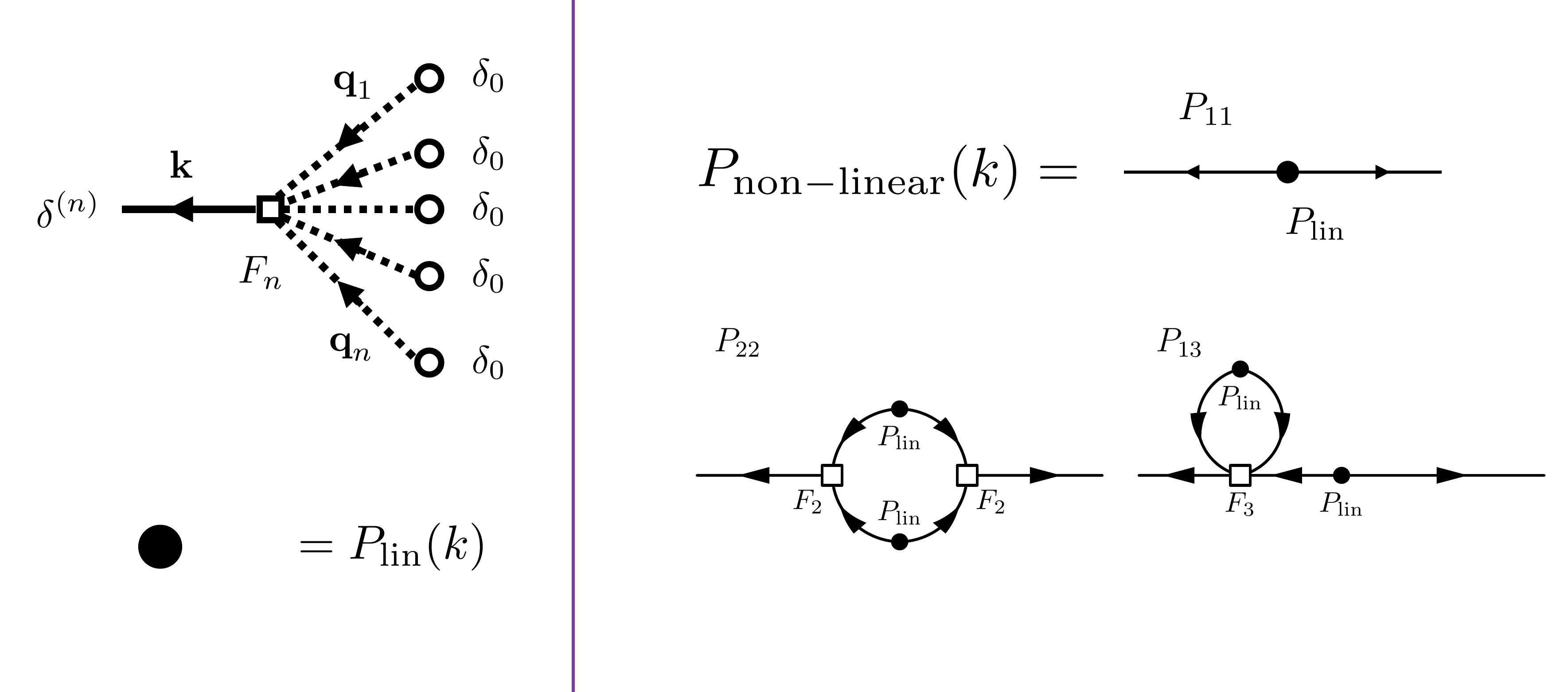}
\caption{
Diagrammatic representation of the $\delta^{(n)}$
correction to the non-linear density field, 
linear matter power spectrum, and the non-linear 
matter power spectrum at the one-loop order.
}
\label{fig:fey}
\end{figure}

Perturbation theory over $\delta^{(n)}$
and $\theta^{(n)}$
can be represented in terms of Feynman diagrams~\cite{Baldauf:2014qfa,Baldauf:2020bsd}.
For example, the $\delta^{(n)}$
correction is shown in Fig.~\ref{fig:fey}.
The linear matter power spectrum, which 
results from a contraction of two linear density fields 
$\langle \delta_0(\k)\delta_0(\k')\rangle$,
can be represented with a filled circle. Then the total 
one-loop power spectrum will be a sum of tree diagrams shown in Fig.\ref{fig:fey}.

\subsection{IR and UV Singularities}

\begin{figure}
\centering
\includegraphics[width=0.49\textwidth]{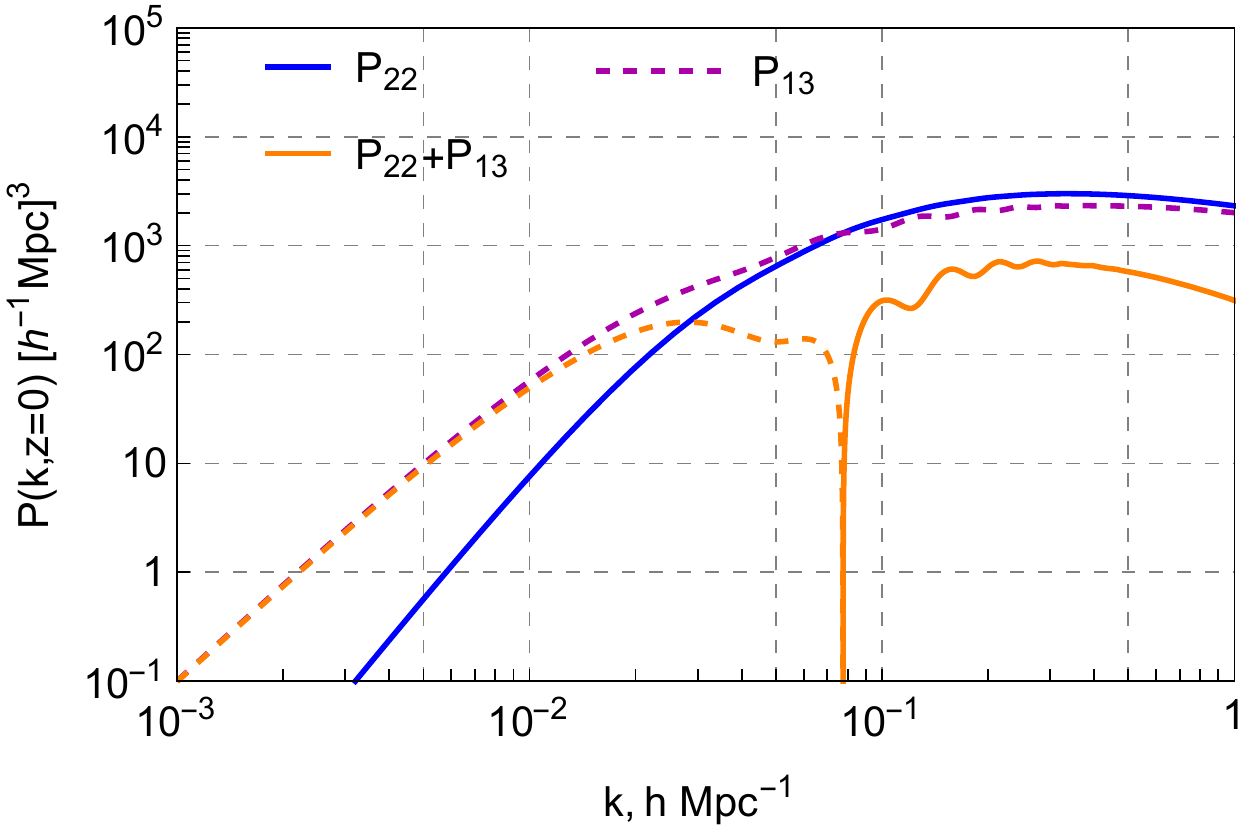}
\includegraphics[width=0.49\textwidth]{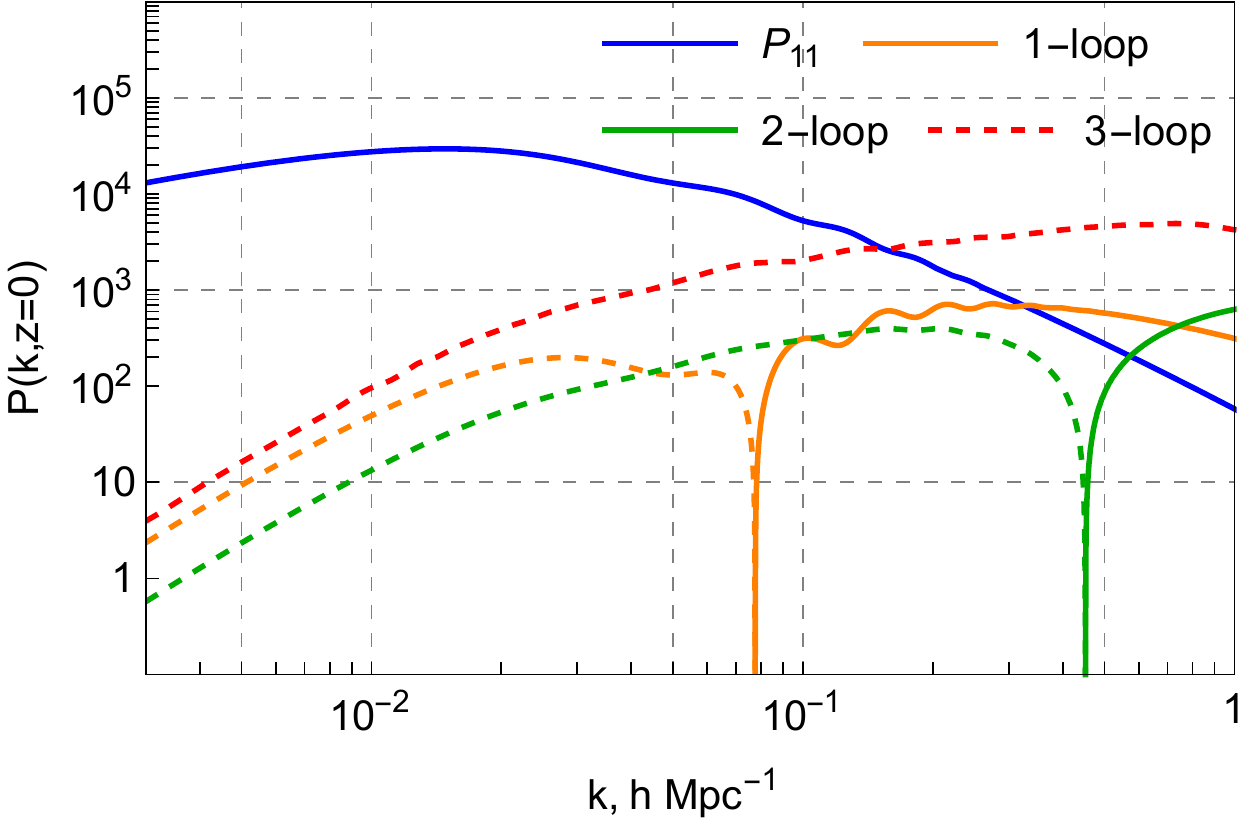}
\caption{\textit{Left panel:}
Individual loop corrections $P_{22}$, $P_{13}$, 
and their sum.
\textit{Right panel:}Tree-level (linear theory), and various loop corrections 
to the matter power spectrum computed within 
the pressureless perfect fluid (PPF) hydrodynamics. 
Dashed curves show negative contributions.
PPF predictions are dominated by displacements that receive
uncontrolled contributions from the UV modes.
This produces a large spurious suppression of the matter power
spectrum on large scales, and leads to a breakdown 
of the PPF perturbation theory.
}
\label{fig:1lbrk}
\end{figure}

Let us consider the IR limits of the one-loop power spectrum integrals. We find that 
the leading IR asymptotics of $P_{13}$ and $P_{22}$ exactly cancel~\cite{Blas:2013bpa},
\be
\begin{split}
& P_{22}(k)\Big|_{q\ll k,~|\k-\q|\ll k}= \frac{k^2}{3}P_{11}(k)  \int_{q\ll k}~\frac{d^3q}{q^2}P_{11}(q) \\
& P_{13}(k)\Big|_{q\ll k}= -\frac{k^2}{3}P_{11}(k) \int_{q\ll k}~\frac{d^3q}{q^2}P_{11}(q)\,.
\end{split} 
\ee
The integrals in the r.h.s. of the above equations 
contain the displacement variance in the IR domain~\eqref{eq:dispvar}. 
Thus, each of these terms describes how the matter power spectrum gets modulated by 
IR displacements. 
The individual corrections $P_{13}$ and $P_{22}$,
which are dominated by their IR limits, 
are quite large, see Fig.~\eqref{fig:1lbrk}.
The total one-loop matter power spectrum that remains
after summing these terms together is five times smaller
than each one-loop term individually.

The fact that individual integrals $P_{13}$ and $P_{22}$
do
converge
is a pure coincidence. 
In fact, the $P_{13}$ and $P_{22}$ integrals could be formally divergent for a power-law spectrum with $P_{11}\propto q^{\nu}$ with $\nu<-1$.
In this respect the IR enhancements are often called ``singularities.''
However, the IR singularities in 
$P_{13}$ and $P_{22}$ for any initial spectrum get exactly canceled once the integrals are summed together.
Although the cancellation is straightforward 
at the one-loop power spectrum level, it becomes much more 
intricate at higher loop order and for higher-order n-point functions,
whose analysis is entangled by subleading IR divergences.
IR singularities appear because the expansion of the final density
field in terms of the initial field 
\eqref{eq:ans_spt} violates the equivalence principle.
The final physical observables, such as the n-point correlation
functions, respect the equivalence principle, which explains 
the cancellations.

\begin{figure}
\centering
\includegraphics[width=0.49\textwidth]{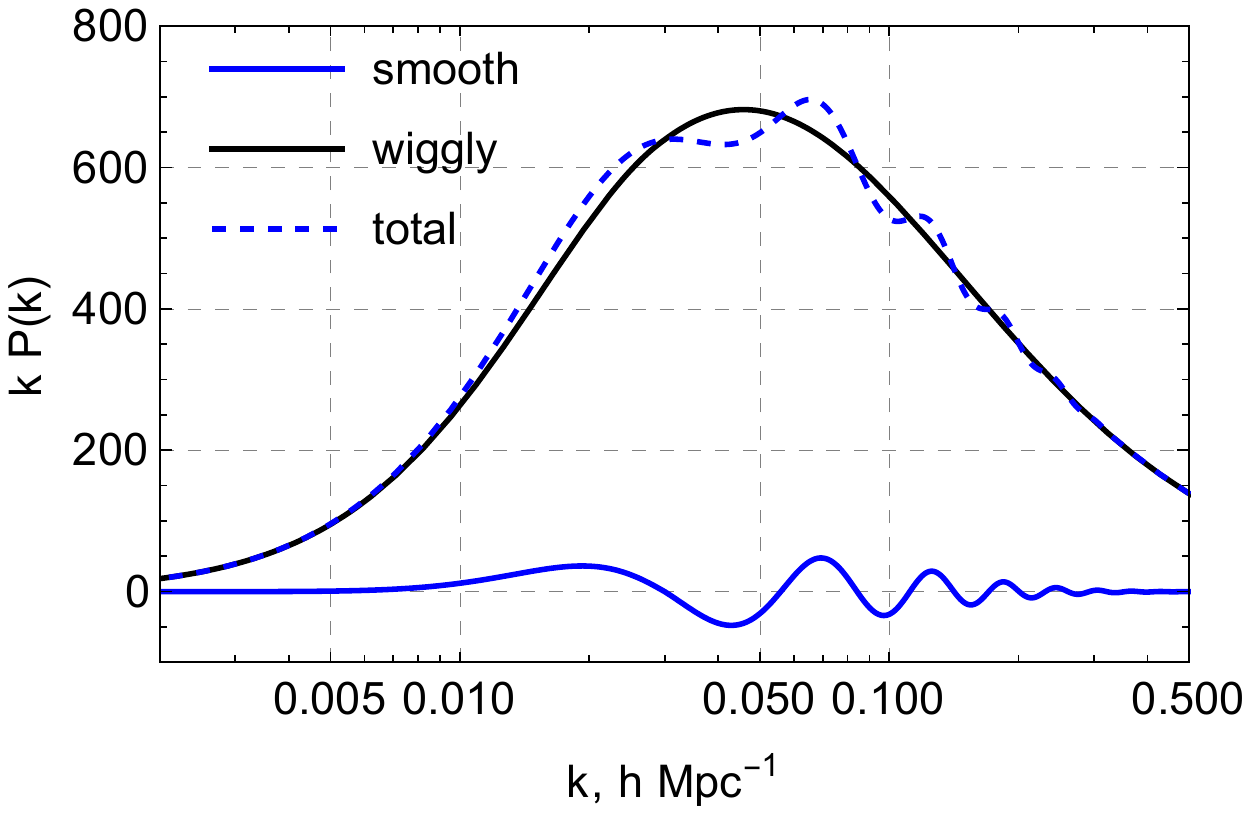}
\includegraphics[width=0.49\textwidth]{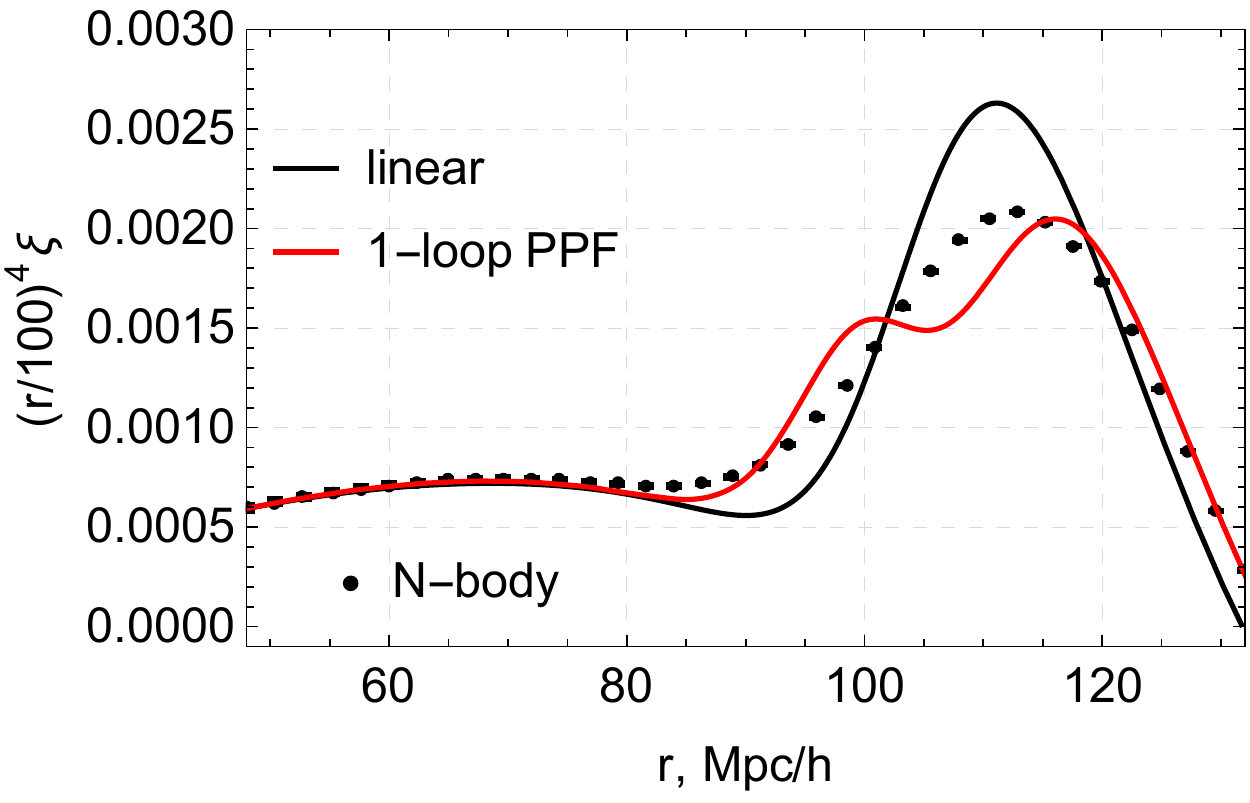}
\caption{\textit{Left panel:}
Wiggly-smooth decomposition of the linear matter power spectrum.
\textit{Right panel:} rescaled $s^4\xi(s)$, where $\xi(s)$ 
is the 
position space two-point correlation
function at $z=0$. Dots are the data measurements
from the Horizon Run N-body simulation.  
Both linear theory and one-loop PPF results
fail to capture the shape of the BAO peak.
}
\label{fig:bao1}
\end{figure}

The non-trivial IR limit obtained after the cancellation 
of the IR singularities is given by~\cite{Blas:2013bpa}:
\be 
\label{eq:1loopir}
P_{{\rm 1-loop}}\Big|_{q\ll k}=\left(\frac{569}{735}-\frac{47}{105}k P'_{11}(k) 
+\frac{1}{10}k^2 P''_{11}(k)
\right)\sigma^2_l(k)\,,
\ee
where $\sigma^2_l(k)\equiv \int_{q\ll k} \frac{d^3q}{(2\pi)^3}P(q)$
is the large-scale mass variance.
We see that the IR modes actually couple the power spectrum through 
large-scale mass fluctuations, and not the displacement. 
Note that we have formally assumed that $P_{11}(|\k-\q|)$ can be Taylor 
expanded around $k$, and its derivatives w.r.t. $k$ are small. 
This is certainly a good assumption if the linear 
power spectrum is a smooth function. 
This is certainly true for $q\ll k\sim 0.1~\hMpc$, see Fig.~\ref{fig:Pk}.
However, $P_{11}$ has a scale associated with 
the baryon acoustic oscillations, $k_{BAO}\sim 10^{-2}~h\text{Mpc}^{-1}$.
For modes with $q\sim k_{BAO}\ll k$ the Taylor expansion above breaks down,
and therefore the nonlinear effects on the BAO need to be resummed~\cite{Senatore:2014via,Baldauf:2015xfa,Blas:2016sfa}. 
To get some intuition on these effects, let us note that the full power 
spectrum can be presented as a sum of the smooth (power-law),
and wiggly components (see Fig.~\ref{fig:bao1}), 
\be
\label{eq:wiggly_smooth}
P_{11}(k)=P_{smooth}(k)+P_{w}(k)\,,
\ee
where the wiggly part can be approximated as  
$P_{w}(k)=f_s(k)\sin(k/k_{\rm BAO})$ ($f_s$ being a smooth envelop). 
Plugging this ansatz into 
\eqref{eq:1loopir}, we obtain the total wiggly power
spectrum at the one-loop order, 
\be
\begin{split}
& P^{{\rm 1-loop}}_w(k)=P_w(k)+\mathcal{O}(1)\times \sigma^2_l P_w(k) -\frac{k}{k_{\rm BAO}}
\left(\frac{47}{105}-\frac{k f'_s}{5 f_s}\right)
-\frac{k^2}{k^2_{\rm BAO}}\frac{\sigma^2_l}{10}P_w(k)\\
&= f_s \underbrace{\left(1-\frac{k^2}{k^2_{\rm BAO}}\frac{\sigma^2_l}{10}\right)}_{\text{BAO smoothing}}
\sin\left(\frac{k}{k_{\rm BAO}}\left\{1-
\underbrace{\left[\frac{47}{105}-\frac{k f'_s}{5 f_s}\right]\sigma^2_l}_{\text{Shift of the BAO phase}}
\right\}\right)+\mathcal{O}(1)\times \sigma^2_l P_w(k)\,.
\end{split}
\ee
We can see two effects here: (a) an overall suppression 
of the amplitude of the BAO (smoothing), 
and (b) the shift of their phase 
w.r.t. linear theory. The effective coupling 
constant $(k/k_{\rm BAO})^2\sigma_l^2$ can be parametrically enhanced for
 $k\gg k_{\rm BAO}$, which can potentially result in $(k/k_{\rm BAO})^2\sigma_l^2$
 being an order one number. This is worrisome, as the two-loop
 correction to $P_w$, which scales like $(k/k_{\rm BAO})^4\sigma_l^4$, 
 is not parametrically suppressed w.r.t. the one-loop result.
 Moreover, we did not take into account the fact that 
 the Taylor expansion of $P_{11}(|\k-\q|)$ inside the $P_{22}$ integral 
 actually breaks down for $q\sim k_{\rm BAO}$, and thus we have 
 a breakdown of naive perturbation theory in two places here.
 This fact is manifest when we compare the predictions of 
 linear theory and one-loop PPF models to the data, see Fig.~\ref{fig:bao1},
 which clearly fail to capture the behavior observed in N-body simulations.
 However, we will see later that there is a way to resumm these kinds of corrections 
 from the perturbative series. This procedure is called IR resummation.

Now let us focus on the 
UV limit of loop integrals, $q\gg k$. We have~\cite{Baldauf:2014qfa}
\be
\label{eq:UVlims}
\begin{split}
& P_{22}(k)\Big|_{q\ll k}= \frac{9}{196 \pi^2 } k^4  \int_{k\ll q}~\frac{q^2 dq}{q^4}P^2_{11}(q) \\
& P_{13}(k)\Big|_{q\ll k}= -\frac{61}{630 \pi^2} k^2 P_{11}(k) \int_{k\ll q}~\frac{q^2dq}{q^2}P_{11}(q)\,.
\end{split} 
\ee
We see that the UV limit of $P_{22}$ does not involve 
$P_{11}(k)$, and moreover it scales like $k^4$ on large scales. 
In agreement with Peebles' arguments, it describes how two short modes generate power on large scales. 
We will discuss this effect in detail shorty.
In contrast, the UV limit of $P_{13}$ 
describes how $P_{11}(k)$ gets modulated by the short-scale displacements.
Importantly, the $P_{22}~(P_{13})$ integrals are divergent 
in the UV for a power-law spectrum with $\nu\geq 1/2~(\nu\geq -1)$. 
However, 
for the actual $\Lambda$CDM power spectrum they turned out to converge. 
This does not mean that they have converged to the right answer though: 
the modes in the domain $q\gg k$ cannot be described by a perfect fluid
at all. In particular, we do not expect to capture the effects of halo 
virialization. 

At the one-loop level the mistake introduced by doing the loop
integrals all the way up to infinity turned out to be small for the 
$\Lambda$CDM power spectrum. However, at higher loop orders it becomes very significant~\cite{Blas:2013bpa,Blas:2013aba}. 
In Fig.~\ref{fig:1lbrk} we show the one-, two-, and three-loop corrections 
to the matter power spectrum in PPFM. We see that the two-loop correction is as big as the one loop,
and the three loop term is larger than the first two loops combined. 
This apparent breakdown of perturbation theory is produced by the 
uncontrolled UV corrections that become more and more sizable at higher loop orders. Note that the leading UV corrections in the low-$k$
limit are negative and proportional to $k^2P_{11}$. Just like in the one-loop
case, they correspond to 
UV displacements. The appearance of these UV enhancements
signals that perturbation theory misses the halo
virialization physics, which would confine the DM particles inside the halo
and reduce their actual displacement in comparison with the perturbation 
theory prediction. 
Indeed, in perturbation theory the two DM particles would continue 
to move to ever increasing distances from one another instead of sticking together 
and forming a halo as shown in
cartoon~\ref{fig:cartoon}.

A more meaningful approach to the UV problem 
would be to introduce a UV cutoff $\Lambda\sim {O}(R^{-1}_{\rm vir})$ in the integrals. But then the answer would necessarily depend on that cutoff. The naive pressureless perfect fluid theory 
does not have any means
to systematically remove the cutoff dependence from the answer.
This problem will be resolved in the EFTofLSS by means of 
a non-trivial stress-tensor.

\subsection{Physical IR effects}

\begin{figure}
\centering
\includegraphics[width=0.69\textwidth]{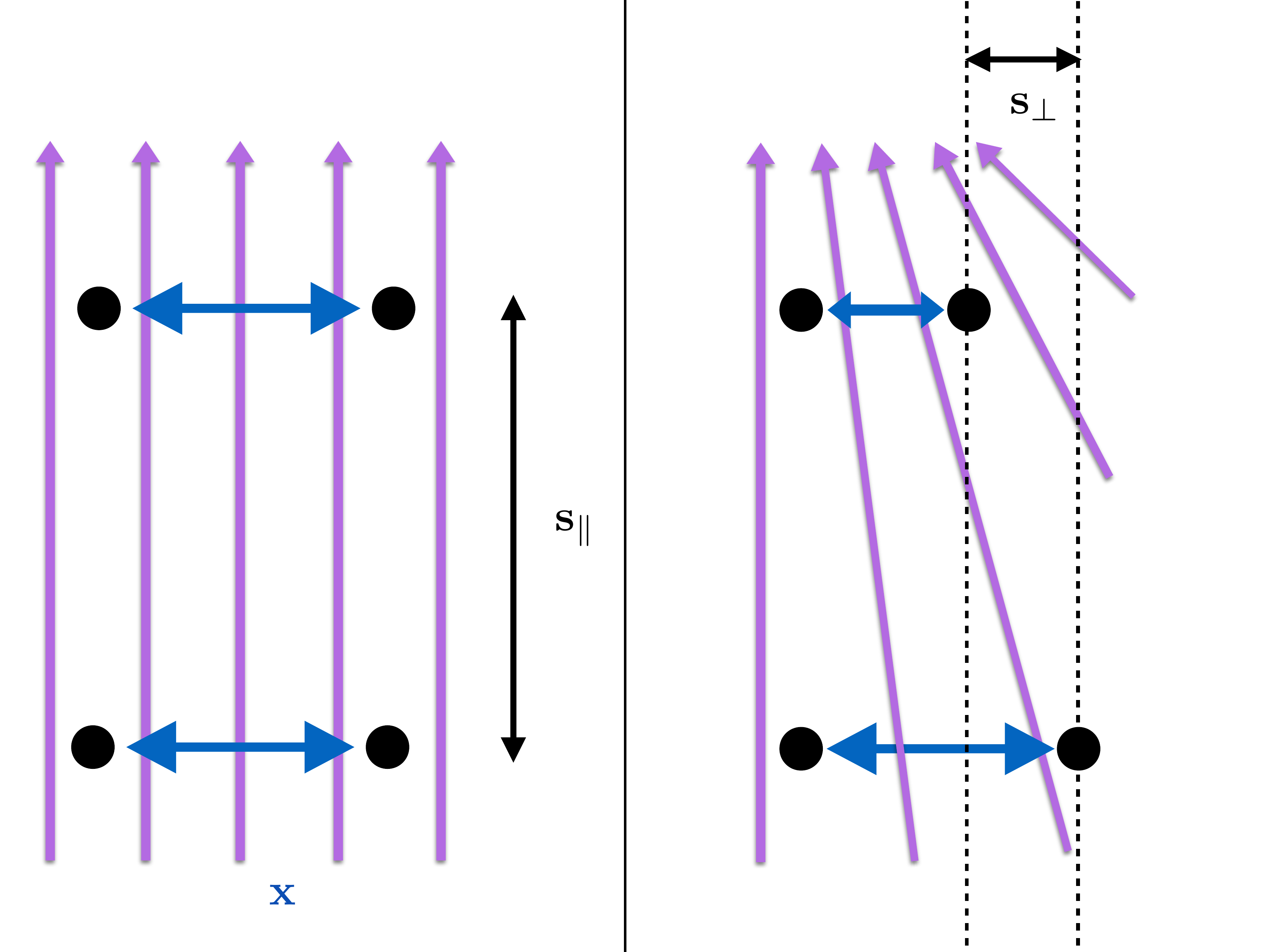}
\caption{
Cartoon picture of non-linear IR effects. \textit{Left panel:}
the homogeneous flow (produced by infinitely soft modes),
does not affect correlation properties of DM particles. 
\textit{Right panel}: if the flow converges (or diverges), 
DM particles would be moved closer (or further). 
Thus, the 2-point correlation function of the DM density
will waken when averaged over many flows.
}
\label{fig:flow}
\end{figure}

Let us discuss the physical IR and UV effects that our perturbation
theory is missing. As far as IR effects are concerned, the problem is 
that these effects happened to be large and need to be treated 
non-perturbatively. Let us get more insight into their dynamics. 
Imagine two dark matter particles in a homogeneous flow.
This flow represents an IR limit of the loop integrals.
The flow can displace each particle by as much as $\sim 6~h^{-1}$Mpc, see Eq.~\eqref{eq:displvar}. 
However, the correlation between 
the two DM particles will not be affected, 
as both of them are displaced by the flow
equally. 
This result can be traced back to the equivalence principle. 
One can think of these two DM particles as being placed in 
an ``Einstein elevator.'' All physical observables, such as the 2-point 
correlation function, should be insensitive to the large scale
acceleration. Cancellation of the effect due to an IR flow thanks 
to the equivalence principle precisely translates into the cancellation
between the $P_{13}$ and $P_{22}$ integrals: each term separately 
describes how the power spectrum is modulated by the 
large scale displacements, but the net effect of such 
displacements should be zero.

To get a rough picture of the effect, let us assume 
that the large-scale (IR) displacement is constant. 
When the two DM 
particles are displaced by a homogeneous displacement field $\s_\parallel$ (see the left panel of Fig.~\ref{fig:flow}),
their correlation function will not change,
\be
\langle \delta(0+\s_\parallel)\delta(\x+\s_\parallel) \rangle  = 
\langle \delta(0)\delta(\x) \rangle \,.
\ee
The above result is a simple consequence of 
the fact that the 
correlation function depends only on the absolute separation between 
the points (statistical homogeneity). 

Now let us imagine that the flow is not 
exactly homogeneous, i.e. the gravitational acceleration
has a gradient. 
This situation is illustrated in the right panel of Fig.~\ref{fig:flow}.
 Physically, this would mean that the two DM particles (or density clumps)
would be displaced from one another. Once we average over 
different patches of space, this should result in a loss of correlation 
between the two particles, and hence the suppression
of the correlation function. Indeed, 
this gradient can be described 
by an additional displacement $\s_\perp$,
\be
\langle \delta(0)\delta(\x+\s_\perp) \rangle  = 
\int_\k e^{i \k \cdot \x}\langle e^{i \k \s_\perp}\rangle P(k) 
=  \langle e^{(\s_\perp\cdot \nabla)}\rangle\int_\k e^{i \k \cdot \x} P(k)\,,
\ee
where we assumed for simplicity 
that the displacement gradient is not correlated with the original 
density field. Using the cumulant expansion theorem we obtain
\be
\label{eq:sperp}
\langle \delta(0)\delta(\x+\s_\perp) \rangle  = 
e^{\sum_{n=1} \frac{1}{n!} \langle (\s_\perp\cdot\nabla)^n_c \rangle } 
\langle \delta(0)\delta(\x) \rangle = e^{\frac{1}{2}\langle s^i_\perp s^j_\perp \rangle \d_i \d_j } \langle \delta(0)\delta(\x) \rangle  \,,
\ee
where in the second equality above we made use that the bulk flows are described by linear theory, i.e. only the second order
cumulant of $\s_\perp$ survives.  
We see that the relevant expansion parameter here is the 
gradient of the correlation function times the displacement.
If the correlation function were exact power-law, $\xi(r)\propto (r/r_{\rm NL})^\gamma$, the bulk flow effects would be perturbative, $\langle s^is_i\rangle /r^2\ll 1$.
Indeed, we could Taylor expand the exponent in \eqref{eq:sperp}
and this expansion would be controlled by a small parameter $(r_{NL}/r)^2$ (we used that in this example $\langle s^is_i\rangle\sim r_{\rm NL}^2$).

However, the effect of bulk flows is non-perturbative for 
the BAO part of the correlation functions, $\xi_{\rm BAO}(r)\propto \exp\{-(2\sigma^2_{\rm BAO})^{-1}{(r-r_{\rm BAO})^2}\}$, where $\sigma$ is the linear width of the BAO peak in position 
space $\sim 5$ $h^{-1}$Mpc. Indeed, 
\be 
\langle s_\perp^2 \rangle \d_r^2 \xi_{\rm BAO}\sim \frac{\langle s^i s_i\rangle }{\sigma^2_{\rm BAO}} \xi_{\rm BAO}\sim \mathcal{O}(1)\times \xi_{\rm BAO}\,,
\ee
i.e. we cannot Taylor expand the exponent in \eqref{eq:sperp}. 
In actual perturbation theory however, 
$\s_\perp \propto \frac{\nabla}{\Delta} \delta^{(1)}$, and thus this 
exponent would only  
show up order by order in the perturbative expansion. 
Note that in this picture the non-pertubative nature of the non-linear evolution
of the BAO results from a similarity between the width of the BAO peak and the 
displacement rms.
As a result of these large displacements, the BAO correlation degrades, 
i.e. the amplitude of the BAO peak diminishes~\cite{Crocce:2007dt}. This coincides with the 
qualitative prediction of the one-loop correction that we have seen earlier.
This behavior is also confirmed 
in N-body simulations, see Fig.~\ref{fig:bao1}.

\subsection{Physical UV effects}

Let us now discuss typical corrections due to 
physics that we are missing
in our perfect fluid toy model. 
Let us estimate typical
corrections due to virialized motions~\cite{Peebles}. 
Our discussion will be based on the famous Peebles' argument for the estimation of the large-scale power spectrum due to virial motions. 
Imagine a distribution of N dark matter particles. Their density in Fourier
space is given by
\be
\rho_\k = \sum_{n} m_n e^{i\k \x_n}\,.
\ee
Now imagine that there is a short scale process 
that moves each particle by a small displacement $\Delta \x_n$. 
This generates the following correction to the density field,
\be 
\begin{split}
\rho_\k =\sum_{n} m_n e^{i\k (\x_n+\Delta \x_n)} = \sum_{n} m_n e^{i\k \x_n}(1+i\k \Delta\x_n -\frac{k_ik_j}{2}\Delta x^i_n \Delta x^j_n +\mathcal{O}(k^3 \Delta x^3))\,.
\end{split}
\ee
The first term in the l.h.s. is the usual background density, the second term 
vanishes due to momentum conservation $(\sum_n m_n \Delta \x_n=0)$, so the first
non-trivial correction due to particles' reshuffling is given by 
\be
\Delta \rho_\k =  
 -\frac{k_ik_j}{2}\sum_{n} m_n e^{i\k \x_n}\Delta x^i_n \Delta x^j_n\,.
\ee
Wee see that the mass and momentum
conservation dictate
the $k^2$-dependence of the perturbed density field.
The typical displacements inside the halo are of the order of the halo
virial radius, $\Delta x_n\sim R_{\rm vir}$, and hence the typical 
corrections to the density field due to the halo 
virial scale as $(R_{\rm vir}k)^2$.
If the short scale 
displacements are completely uncorrelated with large scale modes, 
the total correction to the matter power spectrum can be estimated as
\be 
P_{\rm vir}(k)\sim (k R_{\rm vir})^4  R_{\rm vir}^3 \,.
\ee
This is the famous Peebles' result on the leading behavior of the matter power spectrum generated by
virialized structures. Since this contribution is uncorrelated with long-wavelength modes, it is purely stochastic. 
However, it is also possible that the short scale displacements 
$\Delta x_n$ are actually correlated with the long-wavelength density field $\delta_\k$, i.e. $\delta_{\rm vir}(\k)\propto k^2 R_{\rm vir}^2 \delta_\k$.
In practice, this correlation can be generated by tidal forces~\cite{Baumann:2010tm}.
Then the short scale motions would modulate the matter power spectrum 
on large scales,
\be 
\langle \delta_{\rm vir}(\k')\delta^{(1)}_\k \rangle' = - k^2 R_{\rm vir}^2 P_{11}(k)\,.
\ee
This is the leading deterministic effect of halo virialization 
on the large-scale matter power spectrum. 
A similar effect 
can be generated by pressure. Indeed,
adding the usual pressure term $c_s^2 \nabla \delta \rho /\rho$ to the r.h.s. of 
the Euler equation~\eqref{eq:hydro0},
we get:
\be 
\label{eq:press}
\begin{split}
& \frac{\d}{\d t}v^i + H v^i 
+  v^j  \d_j v^i 
 = -  \d_i \Phi -  c^2_s \d_i \delta \,,
\end{split}
\ee
which gives us the following Euler equation in Fourier space
\be
\dot \Theta_\k -\frac{3}{2}\delta_\k+\frac{1}{2}\Theta_\k = \int_{\q_1\q_2}
\delta^{(3)}_D(\k-\q_{12})
\frac{\q^2_{12} (\q_1\cdot \q_2)}{2q_1^2q_2^2}\Theta_{\q_1}\Theta_{\q_2} - \frac{k^2c_s^2}{\mathcal{H}^2} \delta_\k \,.
\ee
Treating the sound speed term as a perturbation and
using the density Green's function~\eqref{eq:green} we find 
\be
\label{eq:cs0}
\delta_\k^{(c_s^2)} =
\int da' G_\delta(a,a')(-k^2c_s^2 \delta^{(1)}_\k)\equiv -k^2 \gamma \delta^{(1)}_\k\,.
\ee
This produces to the following correction to the matter power spectrum,
\be
P_{\text{1-loop}}(k)= P_{11}(k)+P_{22}(k) + P_{13}(k) - 2 k^2 \gamma P_{11}\,.
\ee
The upshot is that the virialization effects actually 
act like an effective pressure 
in the fluid equations! 

On dimensional grounds the size of the 
pressure parameter $\gamma$ should be of the order of $R_{\rm vir}^2$.
The effective sound speed 
can be estimated as $c_s^2\sim (R_{\rm vir}\mathcal{H})^2\sim 10^{-6}c^2$, where $c$
is the speed of light. 
This is similar to the speed of sound of ordinary water.
We stress however that this effective pressure 
is different from the microscopic pressure of ordinary fluids.
The hydrodynamical description of normal fluids is adequate 
down to scales of order the molecules' mean free path. 
Thus, the 
hydrodynamic description is valid e.g. 
for the dynamics of sound waves. 
In contrast, the
description of the dark matter fluid with 
the effective pressure is valid only for 
$k\ll k_c= 2\pi R^{-1}_{\rm vir}$. 
Since $k_c\sim \mathcal{H}/c_s$ here is the analog of the Jeans' length, the dark matter effective fluid description~\eqref{eq:cs0}
breaks down before it could describe sound waves.

\subsection{Lessons}

Let us summarize the lessons that we have taken from the 
pressureless 
perfect fluid hydrodynamics. First, one needs to work with 
coarse grained fields, as only these fields 
have small fluctuations. Otherwise the fluid description 
does not have a small parameter.

The second lesson is that the 
UV behavior of the theory needs to be properly renormalized. 
In most of physical systems the dynamics of the long-wavelength fluctuations 
must be insensitive to the details of the short scale physics.
This property is called ``decoupling.''
This fact should be reflected in the correct equations for structure formation. 
We have seen that in the PPF formalism this 
is not the case. The loop integrals there
receive large corrections from the short scales where the 
fluid description does not apply. 
And there are no technical 
tools within the PPF to fix this issue and remove these large 
enhancements (``singularities'').
This clearly indicates 
the inconsistency of the PPF approach from the mathematical 
viewpoint.
In addition, there are physical effects of halo virialization
and backreaction on large scales that 
are missing in the PPF hydrodynamics. 

The third important lesson is that there are also IR singularities, 
which plague individual loop contributions. These singularities cancel 
when all corrections are summed together at a given loop order. 
The origin of these singularities is in the fact that the perturbative 
expansion in terms of the density field does not respect the 
equivalence principle. Although this looks like a technical inconvenience, 
the presence of IR singularities obscures the analysis of
physical IR effects that 
alter the BAO feature in the matter distribution.
These effects turned out to be large in our Universe,
and hence require a non-perturbative treatment. 

The first two drawbacks of the PPF approach are straightforward to fix. 
The analysis presented above suggests that this can be done by means of 
using a non-zero 
stress tensor and smoothing the resulting equations of motion.
This is the core idea of the EFTofLSS.
The IR issue, however, is less straightforward. 
The EFT principles itself do not help in this situation because 
the IR singularities are artifacts of the standard Eulerian 
description. 
There is no good solution to this problem in terms of Eulerian hydrodynamics. 
We will show later how this issue is resolved
with another formulation of the EFTofLSS called
time-sliced perturbation theory. 


\section{Large Scale Structure as an Effective Fluid}\label{sec:eftlss}

There are several ways to obtain equations of motions 
for the coarse-grained matter density and velocity fields~\cite{Baumann:2010tm,Carrasco:2012cv}. 
In the ``top-down'' approach
we start with a UV model, e.g. the Newtonian dynamics of N particles. 
This ensemble of particles is described by the Boltzmann-Vlasov equation. 
Coarse graining this equation and taking moments of the 
distribution function we would arrive at 
 equations for a general vicious fluid. 
The process of coarse graining has a
meaning of ``integrating out'' short scale degrees of freedom. 

In the ``bottom-up'' approach 
we start with the most general equations of motion
for large-scale degrees of freedom, which are the DM
density and velocity fields. 
In this case the situation is equivalent to that of
ordinary fluids, with 
the most general description being 
that of an imperfect fluid interacting through gravity.
Then we coarse grain these equations by applying a low-pass filter
with a momentum cutoff $\Lambda$.
The resulting equations will be, again, 
general equations of motion for a viscous fluid.
Thus, from the technical point of view, 
we have to solve now the following system:
\be 
\label{eq:hydro2}
\begin{split}
& \frac{\d}{\d \tau} \delta_\ell + \d_i[(1+\delta_\ell)v^i_\ell] = 0\,,\\
& \frac{\d}{\d \tau}v^i_\ell+ \HH v^i_\ell  
+   v^j_\ell \d_j v^i_\ell 
+  \d_i \Phi_\ell
 = -  \frac{1}{\r_\ell}[\d_j \sigma^{ij}]_\Lambda \,,
\end{split}
\ee
supplemented with the coarse-grained Poisson equation,
\be
\Delta \Phi_\ell = \frac{3}{2}\mathcal{H}^2\Omega_m(\tau) \delta_\ell \,.
\ee
These equations look almost the same as the equation for PPF, 
but however, there are several important differences. 
First, our variables here are long-wavelength 
overdensity and velocity fields~\cite{Carrasco:2012cv,Mercolli:2013bsa}.\footnote{Note that the actual physical variable in our system 
is the coarse-grained momentum field, while the 
velocity 
field in Eq.~\eqref{eq:hydro2} is formally built out of the smoothed momentum and density 
fields. This velocity field is ``bare,'' i.e. 
it should be distinguished from the physical, finite velocity 
field, which differs from $v^j_\ell$ by a counterterm.}
Even though these coarse grained variables explicitly depend 
on the smoothing scale $\Lambda$, the physical observables do not.
Technically, this is realized by means of 
counterterms embedded in the stress tensor.

Second, there is a non-trivial 
stress tensor $\sigma^{ij}$, which encapsulates the effects of 
short-scale backreaction. 
In the EFT approach this backreaction is captured by means of effective 
operators that involve only long distance degrees of freedom such as $\delta_\ell$,
and their gradients. In the context of fluid dynamics 
the most general expression for $\sigma_{ij}$ in a derivative expansion is
\be
\label{eq:NS}
\sigma_{ij} = p\delta_{ij}-\zeta \delta_{ij} (\d_k v^k) -\eta \left(
\d_i v_j
+\d_j v_i
-\frac{2}{3}\delta_{ij}(\d_k v^k)
\right)\,,
\ee 
where $\eta$ and $\zeta$
are shear and bulk viscosities, respectively.
Unfortunately, it turns out that the fluid ansatz~\eqref{eq:NS} 
is not general enough for our purposes.
In particular, we should also add terms involving various 
contractions of the tidal tensor $t_{ij}\propto \d_i \d_j \Phi - \delta_{ij}\Delta \Phi/3$. In addition, we have to go beyond the lowest order 
in the derivative expansion.

Working consistently within EFT approach we should write down 
all possible operators involving long-distance 
degrees of freedom that are consistent with the symmetries 
of the problem. At a given order in perturbation theory there will be only a finite number 
of operators that we need to keep in the theory model.
These operators will appear with free coefficients, which should be treated 
just like Wilson coefficients in other EFTs.
The values of these Wilson coefficients should be determined experimentally.
We stress that they are physical parameters describing our 
cosmological fluid, and neglecting them (or setting them to zero)
is incorrect, just like it is incorrect to neglect bulk and shear viscosities
when describing a fluid. Now we discuss properties of the 
effective stress tensor in detail. 

\subsection{The stress tensor and time-(non)locality}

The quantity that appears in the r.h.s. of the effective 
Euler equation is the smoothed divergence of the effective
stress tensor, $\frac{1}{\r_\ell}[\d_j \sigma^{ij}]_\Lambda$.
This term incorporates the effects of the short scale physics 
and its backreaction on large scales. As we discussed earlier, 
it contains response of short scale (virialized) 
modes to large-scale tidal forces. 
In the EFT approach we need to expand the stress tensor 
in derivatives of the long-wavelength field
while being consistent with rotation symmetry and the equivalence 
principle.
At leading order in the number of density fields 
and its gradients we get\footnote{In principle, one should also add here shear and bulk viscosity 
terms analogous to~\eqref{eq:NS}, but their contribution
is completely degenerate with $\d^i\delta$ 
at the level of the one-loop power spectrum, so we ignore these terms 
for simplicity.}
\be 
\label{eq:stress1}
\frac{1}{\r_\ell}[\d_j \sigma^{ij}]_\Lambda = c_s^2(\tau) \d^i\delta(\x,\tau)\,,
\ee
where $ c_s^2(\tau)$ is a time-dependent Wilson coefficient.
Note that the stress tensor 
cannot depend on the gravitational potential $\Phi$ or its gradient $\d_i\Phi$ -- this would
contradict the equivalence principle, which states that these
quantities can be removed by a gauge choice and hence cannot affect dynamics.

The expression~\eqref{eq:stress1} has a nice property that it
is local in time and space. However, there is no reason why it should be the case. The problem 
is that short wavelength modes, which we integrate out in~\eqref{eq:stress1},  
evolve on the same time scale $\mathcal{H}^{-1}$ as the long wavelength modes.  
Before virialization, the short modes evolve with the same growth factor 
as the long modes.
This means that we cannot really time-average the evolution of short modes. 
Thus, we need to take into account the fact that the short modes affect the entire
evolution of long-wavelenght perturbations along the fluid trajectory. 
We should generalize~\eqref{eq:stress1} to~\cite{Baldauf:2014qfa} 
\be 
\label{eq:stress2}
\frac{1}{\r_\ell}[\d_j \sigma^{ij}]_\Lambda = \int_{\tau_{\rm ini}}^{\tau} d\tau'~K(\tau, \tau') \d^i\delta(\x_{fl}[\x,\tau;\tau'],\tau')\,,
\ee
where $\x_{fl}[\x,\tau;\tau']$ is the trajectory of the fluid element 
with Eulerian coordinate $\x$ at time $\tau$. $\tau'$ parametrizes this trajectory.
$K(\tau, \tau')$ is some (unknown) time-dependent kernel.
This expression is hard to work with. However, we can Taylor-expand 
$\delta(\x_{fl}[\x,\tau;\tau'],\tau')$ around the fluid trajectory, 
\be 
\label{eq:tayflow}
\begin{split}
\delta(\x_{fl}[\x,\tau;\tau'],\tau') 
&=\delta(\x,\tau)+(\tau'-\tau)\frac{D}{D\tau}\delta+
\frac{(\tau'-\tau)^2}{2}\frac{D^2}{D\tau^2}\delta +...\\
\end{split}
\ee
where we introduced the convective derivative $D/D\tau\equiv \d_\tau + v^i\d_i$,
and the peculiar velocity along the flow
$v^i(\x,\tau)\equiv \frac{d\x_{fl}}{d\tau}\Big|_{(\x,\tau)}$.
The integral along the flow can be written 
in terms of Eulerian quantities, 
\be 
\label{eq:stress3}
\begin{split}
&\int_{\tau_{\rm ini}}^{\tau} d\tau'~K(\tau, \tau') \d^i\delta(\x_{fl}[\x,\tau;\tau'],\tau') \\
&= \left[\int_{\tau_{\rm ini}}^{\tau} d\tau'~K(\tau, \tau')\right] \d^i \delta(\x,\tau)+\left[\int_{\tau_{\rm ini}}^{\tau} d\tau'~K(\tau, \tau')(\tau'-\tau)\right] \frac{D}{D\tau}\d^i \delta(\x,\tau) + ...
\end{split}
\ee
Naively this expansion still contains an infinite number of terms, so it does not look
very useful. However, it simplifies
in perturbation theory, where 
we should 
only keep a finite number of fields in convective derivatives.
Indeed, the terms with convective derivatives can be simplified 
by using equations of motion (assuming EdS for simplicity here), 
\be 
\begin{split}
& \frac{D}{D\tau}\delta = \mathcal{H}\Theta(1+\delta)\,,\\
& \frac{D}{D\tau}\Theta = \mathcal{H}^{-1}\left(-\frac{1}{2}
\mathcal{H}^2 \Theta + \frac{3}{2}
\mathcal{H}^2 \delta + \d_iv^j \d_j v^i -\d^i[\rho_\ell^{-1}\d^j \sigma_{ij}]
\right)\,.
\end{split}
\ee
The stress-tensor in the equations above should be computed 
iteratively. From the structure of these equations 
we see that at a given order in perturbation theory 
one has to keep track of a finite number of terms 
in~\eqref{eq:tayflow}. For instance, at one-loop
order it is sufficient to use the linear theory solution
~\eqref{eq:gmode}
in the r.h.s. of~\eqref{eq:tayflow}, in which case 
there will be only one linearly-independent term proportional 
to $\d^i\delta^{(1)}$. Now we can formally sum all the infinite series, which 
can be packaged into an overall coefficient in front of the linear density 
field:
\be
\begin{split}
\frac{1}{\r_\ell}[\d_j \sigma^{ij}]_\Lambda = &
\left[\int_{\tau_{\rm ini}}^{\tau} d\tau'~K(\tau, \tau') 
\left(\prod_{n=0}^\infty \frac{(\tau'-\tau)^n}{n!}\d^n_{\tau}  \right) \right]D(\tau)\d^i\delta_0(\x)\\
=& c_s^2(\tau) \d^i\delta^{(1)}(\x,\tau)\,.
\end{split} 
\ee 
This is the same result as~\eqref{eq:stress1}. 
However, this simplicity does not hold at higher orders 
in perturbation theory, where convective derivatives 
induce operators nonlocal in space.

All in all, non-locality in time is an important property 
of the EFTofLSS. However, it can be removed 
perturbatively, so the resulting expression for the 
stress tensor takes a manifestly local in time expression. 
The time-non locality. however, does not disappear completely:
we trade it for non-locality in space, which 
is produced by convective derivatives as in Eq.~\eqref{eq:tayflow}.

\subsection{Stochasticity}

Small scale perturbations 
may not be statistically correlated over long distances. 
Non-linear collapse of matter and virialization 
processes can erase memory of the initial conditions,
so that the short-scale density field 
becomes partly uncorrelated with the long-wavelenght modes.
In other words, the actual realization of the effective stress tensor 
is expected to be different from its expectation value.
The density field thus develops a stochastic (noise)
component. In order to incorporate this in the EFT we need to 
include a stochastic term in the effective stress-energy tensor, 
\be
 \frac{1}{\r_\ell}[\d_j \sigma^{ij}]_\Lambda \supset J^i\,,\quad \text{where}
 \quad \langle J^i \delta_\ell \rangle = 0\,. 
\ee 
This term propagates a stochastic component to the density field, 
\be
\delta^{\rm stoch.}\propto  \d_i J^i \,.
\ee
Symmetry arguments constrain the scale-dependence of $\delta^{\rm stoch.}$. 
As we have seen above, mass and momentum conservation
dictate that in the $k \to 0$ limit it goes as $\delta^{\rm stoch.}\propto k^2$,
which implies
\be 
\label{eq:Pshoch}
\langle \delta^{\rm stoch.}(\k')  \delta^{\rm stoch.}(\k)\rangle' =P_J(k)=  R_{\rm stoch.}^7 k^4 + ...\,, 
\ee
where $R_{\rm stoch.}$ is some parameter with dimensionality of length.

\subsection{One Loop Power Spectrum in the EFTofLSS}

Knowing the stress tensor at leading order in the EFT, we can now compute 
the full one-loop prediction. Focusing on the scalar part of the EFT Euler 
equation we get:
\be 
\label{eq:hydro0_eft}
\begin{split}
 & \dot \delta_\k -\Theta_\k = \int_{\q_1\q_2}^\Lambda\delta^{(3)}_D(\k-\q_{12})\frac{\q_{12}\cdot \q_1}{q_1^2}\Theta_{\q_1}\delta_{\q_2}\,,\\
& \dot \Theta_\k -\frac{3}{2}\delta_\k+\frac{1}{2}\Theta_\k = \int^\Lambda_{\q_1\q_2}
\delta^{(3)}_D(\k-\q_{12})
\frac{\q^2_{12} (\q_1\cdot \q_2)}{2q_1^2q_2^2}\Theta_{\q_1}\Theta_{\q_2}-\frac{k^2 c_s^2(\tau)}{\mathcal{H}^2}\delta_\k + \frac{\d_i J^i}{\mathcal{H}^2}\,,
\end{split}
\ee
where we have assumed that smoothing out 
the fields with a low-pass filter is equivalent to 
cutting off the integral at $\Lambda$.
An alternative form of this equation is given by Eq.~\eqref{eq:hydro0_f} with a slightly modified source, 
\be
\HH^2\mathcal{S}_\beta \to \HH^2\mathcal{S}_\beta + \tau_{\theta}\,,
\quad \tau_\theta \equiv  \d^i 
\left[\frac{1}{ \rho_\ell}\d^j\sigma_{ij}\right] = 
 \left(c_s^2 \d^2\delta +\d^iJ_i \right)\,.
\ee
This representation is useful because it allows us to easily
evaluate the corrections to the density field produced 
by the effective stress tensor,
\be 
\delta_\k^{(\sigma)}=\int da' G_\delta(a,a')\tau_\theta(a')\,,
\ee
where $G_\delta(a,a')$ is the Green's function
of Eq.~\eqref{eq:hydro0_f}.
It is convenient to present the total density 
field as 
\be
\delta^{\rm NL}_\k= \delta^{(1)}_\k + \delta^{(2)}_\k+ \delta^{(3)}_\k + \delta^{\rm stress.}_\k + \delta^{\rm stoch.}_\k\,,
\ee
where $\delta^{(n)}_\k$ are the corrections coming from the 
PPF part of the EFT equations (in the absence of the stress tensor), 
while 
$\delta^{\rm stress.}_\k$ and 
$\delta^{\rm stoch.}_\k$ are contributions 
coming from the deterministic and stochastic components 
of the stress tensor, respectively. 
The $\delta^{(n)}_\k$ terms as the same as the ones that we computed 
before. The only difference now is that all fields now have support only 
for $k\leq \Lambda$. 

The field
$\delta^{\rm stoch.}_\k$ only correlates with itself. Its power 
spectrum is given in \eqref{eq:Pshoch}. 
Hence, the only new piece that we have to compute is $\delta^{\rm stress.}_\k$.
Its calculation is identical to the 
computation of the effective pressure correction 
in Eq.~\eqref{eq:press},\eqref{eq:cs0}. 
In perturbation theory we can replace $\delta_\k$
in the effective pressure term with $\delta^{(1)}_\k$,
and using the Green's function~\eqref{eq:green}, get
\be 
\label{eq:cs}
\begin{split}
\delta^{\rm stress.}_\k =- \gamma_\Lambda k^2 \delta^{(1)}_\k
\equiv -\int da' G_\delta(a,a')k^2 c_s^2(a')\delta^{(1)}_\k(a')\,.
\end{split}
\ee
The total matter power spectrum at the one-loop order is given by 
\be\begin{split}
P_{\text{non-linear}}(k) = P_{11}(k,\eta) + P_{22}^\Lambda(k,\eta)+P_{13}^\Lambda(k,\eta)
-2\gamma_\Lambda k^2  P_{11}(k,\eta)+P_J(k) \,.
 \end{split}
\ee
Let us see how the EFT corrections help renormalize the one-loop power spectrum. 
For this reason it is important to stress that $\gamma_\Lambda$ and $R_{\rm stoch}$
that we had so far are ``bare'' parameters, i.e. the contain both ``infinite,'' or
$\Lambda$-dependent pieces, and the finite ones, which capture the physical UV effects.
We split the loop integral into the $\Lambda$-dependent 
piece we do not trust and a calculable piece that we trust
 because it's evaluated over the modes where we know that the EFT is correct.
The corresponding regions are defined by a wavenumber
$q_{\rm trust}$. 
Let us consider first the $P_{13}$
contribution,
\be
\begin{split}
 P_{13}^\Lambda = & 6P_{11}(k)\int_{\q,~q\leq q_{\rm trust}}~F_3(\k,\q,-\q)P_{11}(q) \\
& +6P_{11}(k)\int_{\q,~ q_{\rm trust}\leq q \leq \Lambda}~F_3(\k,\q,-\q)P_{11}(q)\,.
 \end{split}
\ee
For simplicity we assume that $\Lambda>q_{\rm trust}\gg k$, in which case the second term 
above can be simplified as (we use the UV limit~\eqref{eq:UVlims})
\be
\begin{split}
6P_{11}(k)\int_{\q,~ q_{\rm trust}\leq q \leq \Lambda}~F_3(\k,\q,-\q)P_{11}(q)= 
-\frac{61}{630 \pi^2} k^2 P_{11}(k) \int^{\Lambda}_{q_{\rm trust}}dq~P_{11}(q)\,.
\end{split} 
\ee
This term potentially diverges in the limit $\Lambda \to \infty$.
This divergence, however, can be explicitly canceled by an appropriate 
choice of $\gamma_\Lambda$:
\be
\gamma_\Lambda =  \gamma_{\rm finite} -\frac{61}{1260 \pi^2}  \int^{\Lambda}_{q_{\rm trust}}dq~P_{11}(q)\,.
\ee
Thus, the physical $P_{13}$-like contribution is independent of $\Lambda$:
\be
 P_{13}^\Lambda-2\gamma_\Lambda k^2 P_{11} = -2 \gamma_{\rm finite} k^2 P_{11}+6P_{11}(k)\int_{\q,~q\leq q_{\rm trust}}~F_3(\k,\q,-\q)P_{11}(q)\,,
\ee
which means that the effective sound speed 
renormalizes the $P_{13}$ loop. In practice this means that we can choose any $\Lambda$
in our calculations, even $\Lambda=\infty$. Note that the 
finite counterterm $ \gamma_{\rm finite}$ is not determined by theory - it should be 
fit from the data.

A similar calculation shows that the $P_{22}$ correction is
successfully renormalized by 
the noise term $P_{J}$:
\be 
\begin{split}
& P_{22}^\Lambda + P_{J} =2\int_{\q,~q\leq q_{\rm trust}}[F_2(\k-\q,\q)]^2
P_{11}(q)P_{11}(\k-\q) \\
& + 2\int_{\q,~ q_{\rm trust}< q \leq \Lambda}[F_2(\k-\q,\q)]^2
P_{11}(q)P_{11}(\k-\q)+P_J \\
&= 2\int_{\q,~q\leq q_{\rm trust}}[F_2(\k-\q,\q)]^2 P_{11}(q)P_{11}(\k-\q) +k^4\left[\frac{9}{196 \pi^2 }   \int_{q_{\rm trust}}^\Lambda~\frac{ dq}{q^2}P^2_{11}(q)
+R^7_{\rm stoch}\right]\\
&= 2\int_{\q,~q\leq q_{\rm trust}}[F_2(\k-\q,\q)]^2 P_{11}(q)P_{11}(\k-\q)  
+(R^{\rm finite}_{\rm stoch})^7k^4\,,
\end{split} 
\ee
where we formally defined
\be
R^7_{\rm stoch} = (R^{\rm finite}_{\rm stoch})^7 - \frac{9}{196 \pi^2 }   \int_{q_{\rm trust}}^\Lambda~\frac{ dq}{q^2}P^2_{11}(q)\,.
\ee
The final, renormalized EFT power spectrum at one-loop
order is given by 
\be
 \begin{split}
 & P_{\text{non-linear}}(k) = P_{11}(k,\eta) +2\int_{\q,~q\leq q_{\rm trust}}[F_2(\k-\q,\q)]^2 P_{11}(q)P_{11}(\k-\q) \\
 &+6P_{11}(k)\int_{\q,~q\leq q_{\rm trust}}~F_3(\k,\q,-\q)P_{11}(q)
-2\gamma_{\rm finite} k^2  P_{11}(k,\eta)+(R^{\rm finite}_{\rm stoch})^7k^4\,.
 \end{split}
\ee
This expression depends on two Wilson coefficients 
$\gamma_{\rm finite}$ and $R^{\rm finite}_{\rm stoch}$.
The physical value of $q_{\rm trust}$ should be around $0.3~\hMpc$
for redshifts $z\sim 0.5$.
In practice. it turned out that 
the one-loop integrals are saturated around $k_{\rm eq}$
for the $\Lambda$CDM spectrum. Hence, 
the one-loop result essentially does not depend on $q_{\rm trust}>k_{\rm eq}$,
so we can even choose $q_{\rm trust}=\infty$.

\subsection{Power Counting and the Scaling Universe}

The key aspect of effective field theory is 
power counting, which allows one to estimate the importance
of various operators. In the EFTofLSS
we have two important parameters: the
overdensity field and its gradients. 
In this section we show that in fact, 
both of them effectively depend on just
one small parameter, which is the ratio
of the wavenumber of interest to the physical
non-linear wavenumber, $k/k_{\rm NL}$. 
This dependence is akin to the $E/M$ dependence 
in other EFTs, e.g. in the EFT of quantum gravity $M=M_{\rm P}$
or $M=\Lambda_{\rm QCD}$ in ChPT~\cite{Donoghue:2017pgk}.

Let us assume that the initial power spectrum 
is a power-law as in Eq.~\eqref{eq:power}. Then the dimensionless power spectrum, 
which is related to position space density variance, 
is given by 
\be
\Delta(k)=\frac{k^3}{2\pi^2}P_{\rm lin} (k)= \left(\frac{k}{k_{\rm NL}}\right)^{n+3}\,.
\ee
Since $\Delta \sim \delta^2$, we see that each density field scales like 
\be
\delta \sim  \left(\frac{k}{k_{\rm NL}}\right)^{\frac{n+3}{2}} \ll 1\quad \text{if}\quad k\ll k_{\rm NL}\,.
\ee
In a scaling Universe one can easily
estimate the finite part of the loop integrals, 
\be 
\Delta^2_{\rm L-loop}(k)=\left(\frac{k}{k_{\rm NL}}\right)^{(L+1)(n+3)}\,.
\ee
The sound speed terms and the stochastic contributions can also be easily
estimated, 
\be
 \Delta^2_{c_s^2}(k)=\left(\frac{k}{k_{\rm NL}}\right)^{ n+5}\,,\quad 
  \Delta^2_{J}(k)=\left(\frac{k}{k_{\rm NL}}\right)^7\,.
\ee
Then the total power spectrum is given by:
\be
\begin{split}
 & \Delta^2_{\rm non-linear}=\underbrace{\left(\frac{k}{k_{\rm NL}}\right)^{n+3}}_{\rm linear}
 +\underbrace{\left(\frac{k}{k_{\rm NL}}\right)^{2(n+3)}}_{\rm 1-loop}
 +\underbrace{\left(\frac{k}{k_{\rm NL}}\right)^{5+n}}_{\rm counterterm}
 + \underbrace{\left(\frac{k}{k_{\rm NL}}\right)^{7}}_{\rm stochastic}\,.
 \end{split}
\ee
For our Universe the matter power spectrum in the quasi-linear regime 
can be approximated as a power law with $n=-1.5$ and $k_{\rm NL}\sim 0.3~\hMpc$ (at $z=0$).
Then we find for $k\sim 0.1$:
\be
\begin{split}
 & \Delta^2_{\rm  linear}<\Delta^2_{\rm  one-loop}<\Delta^2_{\rm  counterterm}\ll 
 \Delta^2_{\rm two-loop}<\Delta^2_{\rm  stochastic}\,.
 \end{split}
\ee
Thus, the stochastic term can be neglected at the one-loop order.
The actual $\Lambda$CDM universe is, of course, not a scaling 
universe. But the scaling universe estimates happen to be quite 
accurate, which makes them useful to get an idea of
the size of higher order corrections in the EFT.

\section{Path Integral Methods. Time-Sliced Perturbation Theory}\label{sec:tspt}

Large-scale structure theory has two key ingredients: initial conditions
and time evolution. The typical physical observables are $n$-point correlation
functions of random stochastic weakly non-Gaussian fields at a given time
of the observation. 
This problem
can be naturally formulated in the language of generating functionals.
In particular, correlation functions in the EFTofLSS can be 
calculated using the following generating functional,
\be
\label{eq:pf0}
Z[J_\delta]=\int \mathcal{D}\delta_0 \mathcal{P}_0[\delta_0]e^{\int_\k J_\delta(-\k)\delta_\k[\delta_0]} \,,\quad 
\text{where}\quad \mathcal{P}_0[\delta_0]=\mathcal{N}^{-1}\exp\left\{\int_\k \frac{|\delta_0(\k)|^2}{2P_{\rm lin}(k)}\right\}\,,
\ee
$\mathcal{N}^{-1}$ is a normalization constant,
and the functional $\delta[\delta_0]$ is obtained though a perturbative solution 
of the fluid equations, $\delta[\delta_0]=D_+\delta_0+D_+^2[F_2\delta_0]^2+\cdots$.
This relationship encodes the dynamics of the system. 
The path integral above is over all possible configurations of the 
initial density field. 
This field is assumed to have a Gaussian probability density functional (PDF) 
$\mathcal{P}_0$. The n-point correlator 
can be obtained by taking n derivatives of the generating functional w.r.t.
the source $J_\delta$,
\be
\begin{split}
\langle \delta_{\k_1}...\delta_{\k_n}\rangle
&=\frac{\delta^{n}Z[J_\delta]}{\delta J_\delta(-\k_1)...\delta J_\delta(-\k_n)}\Bigg|_{J_\delta=0}=\int \mathcal{D}\delta_0 ~\left(\prod_{i=1}^n\delta(\k_i)\right)\mathcal{P}_0[\delta_0] \,.
\end{split}
\ee
The initial field, however, is not a directly observable physical quantity.
If we want to work in terms of 
actual observables at a final time, it is natural to equivalently rewrite the 
generating functional~\eqref{eq:pf0} as
\be
\label{eq:pf1}
Z[J_\delta]= \int \mathcal{D}\delta  \mathcal{P}[\delta; \tau]
e^{\int_\k J_\delta(-\k)\delta_\k }
\,, 
\ee
where the integration runs over configurations of the 
final density field $\delta$. Its probability 
distribution function, however, $\mathcal{P}[\delta; \tau]$
is not Gaussian anymore. 
Note that we also do not need to know the functional $\delta[\delta_0]$.
Indeed, in the expression~\eqref{eq:pf1}
we have traded the time-evolution of fields for the time-evolution 
of their PDF. And it is the cumulants of this PDF that we
are ultimately interested in. 

The approach in which one studies the PDF of large-scale structure
at a finite time 
instead of cosmological random fields is called 
\textit{time-sliced perturbation theory}~(TSPT)~\cite{Blas:2015qsi,Blas:2016sfa,Ivanov:2018lcg,Vasudevan:2019ewf}.
TSPT is a tool for efficient calculations of cosmological
equal-time correlation functions.
In this 
section we give a brief overview of this method 
and emphasize its advantages over the standard EFTofLSS 
formulation for IR resummation and UV renormalization.

\subsection{Generating Functional for Large-Scale Structure}

In cosmological perturbation theory we are interested 
in overdensity and velocity divergence fields. 
For adiabatic initial conditions these two fields are 
fully correlated, see~\eqref{eq:gmode}. 
It is convenient to chose the velocity divergence field 
as a statistically independent variable in 
the generating functional. We will denote its PDF 
as $\mathcal{P}[\Theta,\eta]$. 
For simplicity, let us assume that the dynamics 
is described by the PPF equations. This will be sufficient
for the purposes of this Section. Then, at any moment of time 
the overdensity field can be expressed through $\Theta$ as
\be
\label{eq:denstt}
 \delta(\k) \equiv \delta[\Theta;\eta,\k] = \sum_{n=1}^\infty \frac{1}{n!}\int_{\q_1}...\int_{\q_n}K_n(\eta;\q_1,...,\q_n)(2\pi)^3\delta_D^{(3)}(\k-\q_{1...n})
 \prod_{j=1}^n\Theta_{\q_j}\,.
\ee
Note that in contrast to the PPF ansatz~\eqref{eq:ans_spt}, the r.h.s. of this 
equation features the fully non-linear field $\Theta$. 
The expansion~\eqref{eq:denstt} can be inserted into the Euler equation~\eqref{eq:hydro0_f} to obtain the following equation of motion for the velocity
field,
\be
 \d_\eta \Theta(\k) \equiv \mathcal{I}[\Theta;\eta]= \sum_{n=1}^\infty \int_{\q_1}...\int_{\q_n}\frac{1}{n!}I_n(\eta;\q_1,...,\q_n)\delta_D^{(3)}(\k-\q_{1...n})\prod_{j=1}^n\Theta_{\q_j}\,.
\ee
Note that in the EdS approximation the kernels $K_n$ and $I_n$ are time-independent.
$I_1=K_1=1$ corresponds to the standard adiabatic growing mode. 

The generating functional for correlation functions of $\delta$ 
and $\Theta$
is given by 
\be 
Z[J_\delta,J;\eta]=\int [\mathcal{D}\Theta]\mathcal{P}[\Theta;\eta]
\exp\left\{ \int_\k \Theta(\k)J(-\k) +\int_\k \delta[\Theta;\eta,\k]J_\delta(-\k) \right\}\,.
\ee
Equal-time correlation functions of $\delta$ and $\Theta$ are
obtained by taking functional derivatives w.r.t. sources
$J_\delta$ and $J$, e.g. the matter power spectrum 
is given by 
\be
(2\pi)^3 \delta_D^{(3)}(\k+\k')P(\eta;k)=\frac{\delta^2Z}{\delta J_\delta(-\k)\delta J_\delta(-\k')}\Bigg|_{J_\delta=J=0}\,.
\ee
The conservation of probability implies the Liouville equation for the PDF:
\be 
\label{eq:liu}
\partial_\eta \mathcal{P}[\Theta,\tau] +\int_\k \frac{\delta}{\delta\Theta(\k)} (\mathcal{P} [\Theta,\tau]\mathcal{I}[\Theta;\eta]) = 0\,.
\ee
By analogy with QFT, we can rewrite the PDF 
perturbatively in $\Theta$,
\be
\label{eq:genfGn}
\begin{split}
\mathcal{P}[\Theta,\tau] =  \mathcal{N}^{-1}\exp\left\{-\sum_{n=1}^\infty \frac{1}{n!}\int_{\k_1}...\int_{\k_n}
\Gamma_n^{\rm tot}(\eta;\k_1,...,\k_n)\prod_{j=1}^n \Theta(\k_j)\right\} \,,
\end{split}
\ee 
where $\mathcal{N}$ is a normalization constant. 
The expression above is reminiscent of the 1PI QFT effective action.
Plugging this into the 
Liouville equation we obtain the hierarchy of equations for 
vertices $\Gamma_n^{\rm tot}$,
\be
\label{eq:liupert}
\begin{split}
& \dot\Gamma_n^{\rm tot}(\eta;\k_1,...,\k_n) \\
& +\sum_{m=1}^n\sum_{\sigma}\frac{I_m(\eta;\k_{\sigma(1)},...,\k_{\sigma(m)})}{m!(n-m)!}
\Gamma^{\rm tot}_{n-m+1}\left(\eta; \sum_{i=1}^m \k_{\sigma(i)},\k_{\sigma(m+1)},...,\k_{\sigma(n)}\right)\\
&=(2\pi)^3\delta_D^{(3)}(\k_{1....n})\int_\q I_{n+1}(\eta;\q,\k_1,...,\k_n)\,,
\end{split} 
\ee
where the sum in the second term on the l.h.s. 
is done over all permutations $\sigma$ of $n$
indices. It is useful to decompose the solution of this equation as,
\be
 \Gamma_n^{\rm tot} = \Gamma_n +C_n\,,
\ee
where $\Gamma_n $ is the solution of the homogeneous 
equation with the initial conditions matching the initial
statistical distribution, whilst $C_n$
is the solution of the nonhomogeneous equation with 
vanishing initial conditions. 
Then $\Gamma_n$ have the meaning of the 
1 particle irreducible (1PI) contributions to the 
tree-level equal time n-point functions 
with ``amputated'' external propagators. 
In contrast, $C_n$ have the meaning of counterterms
that cancel ultraviolet divergences in loop integrals. 

In order to solve Eq.~\eqref{eq:liupert}
we adopt the EdS approximation for kernels $I_n$, in which case
we can use an ansatz that separates time and momentum
dependence, 
\be
\Gamma_n (\eta;\k_1,...,\k_n) = \sum_{l=2}^n e^{-l\eta}\Gamma_n^{(l)}(\k_1,...,\k_n)\,,\quad n\geq 2. 
\ee
Note that $\Gamma_1=0$ as a result of statistical 
homogeneity, $\langle \Theta \rangle = 0$.
This implies the following recursion relation for $\Gamma_n^{(l)}$ with $l<n$,
\be
\begin{split}
&\G_n^{(l)}(\k_1,...,\k_n)=\\
&-\frac{1}{n-l}\sum_{m=2}^{n-l+1}\sum_{\sigma} 
\frac{I_m(\k_{\sigma(1)},...,\k_{\sigma(m)})}{m!(n-m)!}
\Gamma_{n-m+1}\left(\eta; \sum_{i=1}^m \k_{\sigma(i)},\k_{\sigma(m+1)},...,\k_{\sigma(n)}\right)\,.
\end{split}
\ee
Note that all $\G_n^{(l)}$ with $l<n$ are uniquely determined by 
vertices of lower orders. The vertex $\Gamma_n^{(n)}$
should be fixed by initial conditions. Without loss of generality, 
we assume that they are set at $\eta \to -\infty$, and
\be
\lim_{\eta \to -\infty } e^{n\eta}\Gamma(\eta;\k_1,...,\k_n)=\Gamma_n^{(n)}(\k_1,...,\k_n)\,.
\ee
One can derive similar equations for the counterterms $C_n$
and find that they are time-independent and completely fixed by
the kernels $I_n$. 

\begin{figure}
\includegraphics[width=0.99\textwidth]{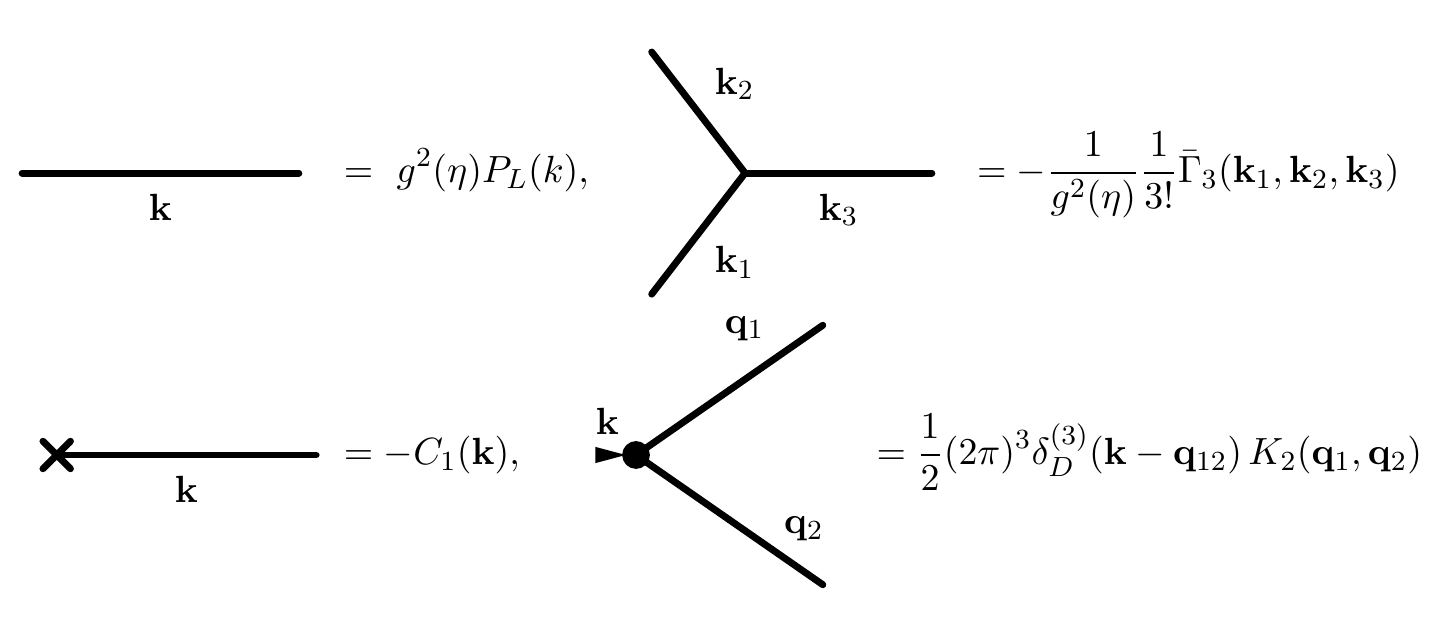}
\centering
\caption{
Examples of TSPT Feynman rules. 
}
\label{fig:tspt_fey}
\end{figure}

For the Gaussian initial conditions 
the solution to Eq.~\eqref{eq:liupert} simplifies greatly. 
In this case
the initial distribution is given by 
\be
\lim_{\eta \to -\infty } \mathcal{P}[\Theta;\eta] =  \mathcal{N}^{-1}
\exp\left\{ -\int_\k \frac{|\Theta_\k|^2}{2D_+^2(\eta)P_{\rm lin}(k)}\right\}\,,
\ee
where we introduced
\be
g(\eta)\equiv D_+(\eta) =e^\eta\,.
\ee
This implies 
\be
\G_2^{(2)}=\frac{(2\pi)^3\delta_D^{(3)}(\k_1+\k_2)}{P_{\rm lin}(k_1)}\,,\quad  
\G_n^{(n)}(\k_1,...,\k_n)= 0 \quad \text{for}\quad n>2\,.
\ee
This means that all $\Gamma_n^{(l)}$ with $l>2$ vanish, and the solution
is simply 
\be
\begin{split}
& \G_n(\k_1,...,\k_n)=\frac{1}{g^2(\eta)}\bar\G_n(\k_1,...,\k_n)\,,\\
& \bar\G_n(\k_1,...,\k_n)= -\frac{1}{n-2}\sum_{m=2}^{n-l+1}\sum_{\sigma} 
\frac{I_m(\k_{\sigma(1)},...,\k_{\sigma(m)})}{m!(n-m)!}
\bar\Gamma_{n-m+1}\left(\sum_{i=1}^m \k_{\sigma(i)},\k_{\sigma(m+1)},...,\k_{\sigma(n)}\right)\,.
\end{split} 
\ee
Remarkably, in the case of Gaussian initial conditions 
and in the EdS approximation 
all vertices have a universal time-dependence $\propto g^{-2}$,
and all of them are sourced by the Gaussian initial weight and hence $P_{\rm lin}(k)$.
The parameter $g^2$ plays a role of the expansion 
parameter (coupling constant). Due to momentum conservation
the vertices are proportional to the Dirac $\delta$-functions
so it is convenient to introduce primed vertices where such 
$\delta$-functions are stripped off,
\be
 \bar\G_n(\k_1,...,\k_n) =(2\pi)^3\delta^{(3)}_D(\k_{1...n}) \bar\G'_n(\k_1,...,\k_n)\,.
\ee
Once the tree-level 1PI n-point functions are fixed, 
the calculation of equal-time correlation functions proceed by a
perturbative expansion of the generating functional~\eqref{eq:genfGn}
around the Gaussian weight. This is identical to the perturbative
calculations of n-point functions in QFT. 
Just like in the usual QFT,
this computation can be represented in terms of Feynman diagrams.
These diagrams are built of vertices $\Gamma_n$, $n\geq 3$, 
and lines correspond to propagators $g^2P_{\rm lin}$, see Fig.~\ref{fig:tspt_fey}. 
One should also include vertices corresponding to counterterms $C_n$, 
$n\geq 1$ in order to subtract certain UV divergences in loop diagrams. 
In this sense counterterms appear in TSPT quite naturally. 

To compute an $n$-point correlation function of the velocity divergence
 one has to draw all diagrams with $n$ external legs. 
It is easy to see that diagrams with larger number of loops are proportional to higher powers of $g(\eta)$. 
Hence, $g(\eta)$ plays the same role as a coupling constant in QFT. 
For the correlators of the density field $\delta$ one should 
use the expression~\eqref{eq:denstt} which is akin to an expression for composite operators in QFT. 
It gives rise to additional vertices proportional to the kernels $K_n$; these are denoted by an external arrow, see Fig.~\ref{fig:tspt_fey}.

\subsection{Soft Limits and IR Safety}

Since we have used the PPF equations as sources in the TSPT 
PDF calculations, our resulting expressions
for the equal-time correlation functions 
must be identical to that of the PPF hydrodynamics 
that we discussed before. This is indeed the case.
However, the intermediate calculations required to obtain
n-point functions are completely different. The PPF 
loop diagrams contain unphysical IR divergences
that cancel only when all diagrams of a given order are summed together. 
In contrast, this problem is absent in TSPT, where all loop diagrams 
are manifestly IR safe. This is to be expected as 
the individual building blocks of TSPT are all IR safe
since they correspond to the physical 1PI equal-time correlators.

One can prove that the vertices $\bar\Gamma_n$, $C_n$ and $K_n$
appearing in the TSPT Feynman graphs are bounded at finite values of 
their arguments. If we split the arguments of a $n$-point kernel 
$\bar\Gamma_n$ into 
``hard'' momenta $k_1,...,\k_l$ that we keep fixed, and ``soft''
momenta $\q_1,...\q_{n-l}$ that we send to zero uniformly as $\q_i=\varepsilon \q'_i$, $\varepsilon\to 0$,
we find that it does not have any poles
\be
\label{eq:irsafe}
\lim_{\varepsilon \to 0}  \bar\Gamma_n(\k_1,...,\k_l,\varepsilon\q'_1,...,\varepsilon\q'_{n-l})=\mathcal{O}(\varepsilon^0) \,.
\ee
The same is true for $C_n$ and $K_n$. 
The IR safety of TSPT vertices~\eqref{eq:irsafe}
can be contrasted with the PPF kernels $F_n,G_n$, which have poles
at soft momenta, see~\eqref{eq:poles}.
The absence of IR singularities of TSPT can be shown to be a direct 
consequence 
of the equivalence principle.

\subsection{IR Resummation}

\begin{figure}
\includegraphics[width=0.99\textwidth]{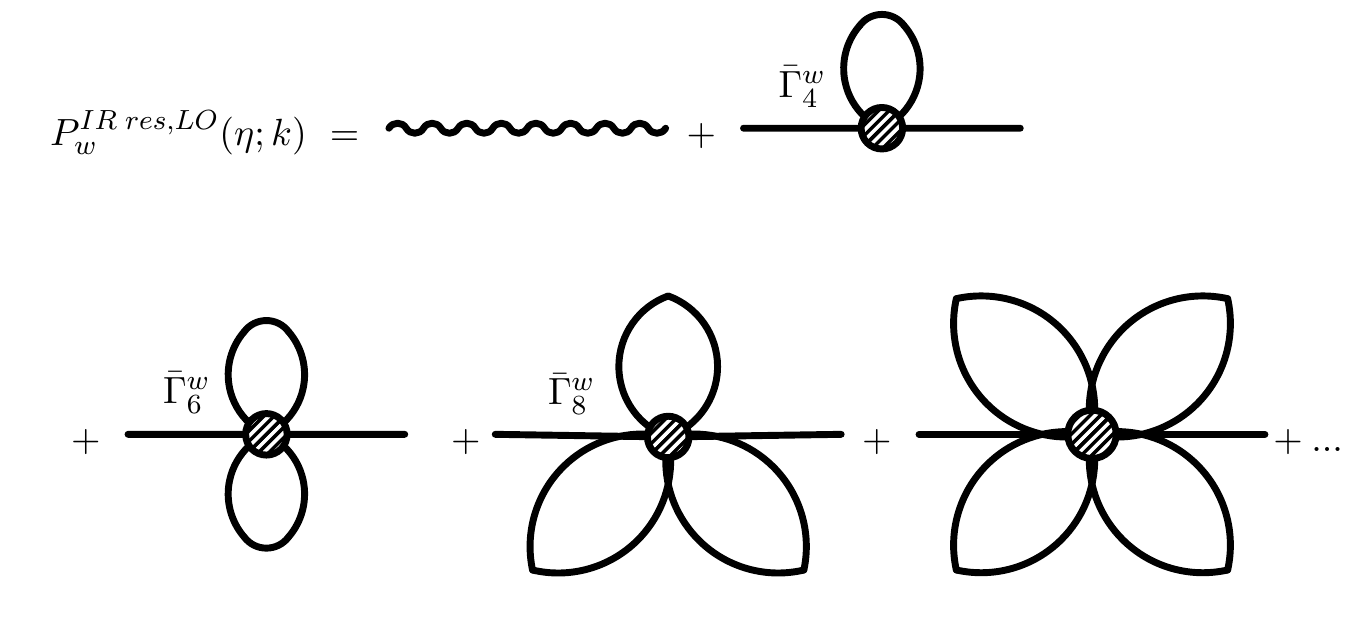}
\centering
\caption{
Feynman diagram representation of IR 
resummation at leading order in the IR enhancement. 
}
\label{fig:daisy}
\end{figure}

The transparent IR structure of the TSPT perturbative expansion
allows us to identify and resumm physical IR contributions
that are responsible for the non-linear evolution 
of the BAO. In order to identify these contributions in 
the perurbative series it is customary to split 
the initial power spectrum into the smooth component $P_{\rm smooth}$ and 
an oscillatory (``wiggly'') contribution $P_w$~\eqref{eq:wiggly_smooth}. 
This decomposition induces a similar split of the 
TSPT vertices as they are sourced by the linear matter power spectrum,
\be
\bar \Gamma_n =  \bar \Gamma_n^s+\bar \Gamma_n^w\,.
\ee
Then, one finds that the wiggly vertices are enhanced in the soft limit. 
For instance, the 3-point vertex expanded for $q\ll k$ and to linear order 
in  $P_w$ is given by 
\be
\label{eq:g3w}
\bar \Gamma_3 \to \frac{\k\cdot \q}{q^2}\left(\frac{P_w(|\k+\q|)-P_w(q)}{P_{\rm smooth}^2(k)}\right) \,.
\ee
In the formal limit $q\to 0$ the difference between the two power spectra
in the numerators vanishes as $q$ and hence cancels the $1/q$ enhancement 
in agreement with the equivalence principle.
However, the Taylor expansion of $P_w(|\k+\q|)$
blows up for $k_{\rm BAO}\lesssim q\ll k$. Non-linear
correlation functions receive large corrections 
from modes in this range. Fortunately, using TSPT one can identify
the enhanced contributions and systematically resumm them to any 
desired accuracy order. The key point is that one can derive
an analog of~\eqref{eq:g3w} for any n-point function and any number of soft momenta. This yields
the following factorization formula:
\be
\label{eq:Gwn1}
 \bar \Gamma'^w_n\left(\k_1,...,\k_m-\sum_{j=1}^{n-m} \q_j,\q_1,...,\q_{n-m}\right)
 =(-1)^{n-m}\left(\prod_{j=1}^{n-m}D_{\q_j}\right)
 \Gamma'^w_m\left(\k_1,...,\k_m\right)\,,
\ee
where $D_\q$ is a differential operator with the following action on the 
wiggly power spectrum,
\be
D_\q P_w(k)=\frac{\k\cdot \q}{q^2}\left(e^{\q\cdot \nabla_{\k'}}-1\right) P_w(k')\Big|_{k'=k}\,.
\ee 
One sees that for $q / k=\varepsilon \ll 1$ each $D_\q \sim \varepsilon^{-1}$, i.e. 
\be 
\label{eq:Gwm}
\bar \Gamma'^w_n\left(\k_1,...,\k_m-\sum_{j=1}^{n-m} \q_j,\q_1,...,\q_{n-m}\right)=\mathcal{O}(\varepsilon^{m-n})\,.
\ee 
This property allows us to formulate power counting rules
that determine the order of enhancement (``degree of IR divergence'') 
of each TSPT diagram. 
For that one has to split each loop momentum into hard and soft 
regions, separated by a scale $k_S$. Then we just have to count the 
number of soft legs attached to
a wiggly vertex $\bar \Gamma^w_n$ of a given TSPT diagram. Eq.~\eqref{eq:Gwm}
tells us that a diagram with the largest number of soft legs will be most enhanced
in the IR. At leading order in the degree of infrared enhancement 
and zeroth order in hard loops
this corresponds to the  ``daisy'' diagrams, see Fig.~\ref{fig:daisy}. 
The daisy diagrams nicely exponentiate,
\be
\begin{split}
&P_w^{\text{IR res, LO}}=e^{-\Sigma^2 k^2 }P_w(\eta;k)\quad \text{where} \\
& \Sigma^2=\frac{4\pi}{3}\int_0^{k_S}dqP_{\rm smooth}(q)
\left[1-j_0\left(qr_{\rm BAO}\right)+2j_2\left(qr_{\rm BAO}\right)\right] \,.
\end{split}
\ee
Note that the expansion in terms of the wiggly-smooth spectra 
is simply a bookkeeping tool that allows one to keep track of the 
enhanced contributions. Since the smooth contributions are never enhanced
in the IR, one can combine the wiggly and smooth parts after resummation
without changing the perturbative order of the resummed corrections.
This amounts to promoting the factorization formula 
Eq.~\eqref{eq:Gwn1} to the total TSPT vertex $\Gamma_n$.
This procedure allows one to circumvent any ambiguity in the 
wiggly-smooth split. In practice, however, this ambiguity
appears only in higher orders of the perturbative expansion
and therefore is negligible for practical applications. 

The key advantage of the TSPT approach for IR resummation 
is that it provides us with a systematic 
program of resumming enhanced IR contributions that can
be carried out to any required accuracy. Thus, it allows one 
to go beyond the leading order in a controlled fashion. 
For instance, one can include hard loops and obtain an expression for the IR-resummed
matter power spectrum at one-loop order,
\be
\label{eq:1loopIRres}
\begin{split}
P^{\rm IR res, NLO}(\eta;k)=&P_s(\eta;k)+e^{-\Sigma^2 k^2}P_w(\eta;k)(1+\Sigma^2 k^2)\\
& P_{\rm 1-loop}[P_s+e^{-\Sigma^2 k^2}P_w]\,,
\end{split} 
\ee
where $P_{\rm 1-loop}[P_{\rm lin}]$ is the 
one-loop integral treated as a functional of the linear 
matter power spectrum. 
The expression~\eqref{eq:1loopIRres} is operationally 
very simple: at one-loop order one just has to compute the 
one-loop integrals using the tree-level IR resummed power
spectrum as an input instead of the usual linear power spectrum.
This simple prescription is reminiscent of the Schwinger-Dyson
resummation approach in non-perturbative QFT. Note however, that 
Eq.~\eqref{eq:1loopIRres} is not a simple ad-hoc prescription, 
it is a result of the rigorous and systematic resummation 
of Feynman diagrams.

Importantly, within TSPT IR resummation can be easily expended 
to higher order n-point functions at an arbitrary order 
in hard loops and for any (subleading) degree of infrared enhancement.

\subsection{UV Renormalization}

Another advantage of TSPT is that it
allows for a formulation of the EFTofLSS in the language of Wilsonian renormalization group within the 3-dimensional Euclidean QFT.
Let us show how these ideas work in practice.

Let us derive the Wilson-Polchinski renormalization 
group equation~\cite{Wilson:1971bg,Polchinski:1983gv} for large-scale structure.
Imagine that we start with a full action for matter clustering 
and integrate out short-scale modes of the velocity divergence 
field $\Theta$ up to momentum scale $\Lambda$. 
This procedure should not affect physical n-point
functions: integrating out the short scales should change the effective
vertices in a way that would keep the low energy physics intact.
This gives rise to the renormalization group (RG) flow of the effective action.
To derive the flow we cut off the free propagator with a sharp
momentum cutoff\footnote{Any other reasonable choice of the cutoff function leads to the same result.},
\be
\label{eq:Pcutoff}
P^{\Lambda}(\eta;k) \equiv P_L(\e;k)H(\Lambda-k)=\Bigg\{
  \begin{tabular}{cc}
  $P_L(\e;k)$, & $k< \Lambda$ \\
  0, & $k> \Lambda$
  \end{tabular}\,,
\ee
Our partition function for large-scale structure then takes the form
\be 
\label{eq:Zpol4}
Z[J]=\int [\mathcal{D}\Theta] \exp\Bigg\{-\int_\p\left(\frac{\Theta(\p)\Theta(-\p)}{2P^\Lambda(\e;p)}+J(-\p)\Theta(\p)\right)+L_{int}(\Theta,\Lambda)\Bigg\}\,.
\ee
Requiring that $Z[J]$ do not depend on the cutoff $\Lambda$,
we obtain the standard Polchinski RG equation \cite{Polchinski:1983gv},
\be 
\label{eq:rg}
\Lambda\frac{\d L_{int}}{\d \Lambda}=-\frac{1}{2}\int_\p\Lambda\frac{\d P^\Lambda(\e;p)}{\d \Lambda}\left(
\frac{\d L_{int}}{\d \Theta(-\textbf{\p})}\frac{\d L_{int}}{\d \Theta(\textbf{\p})}
+\frac{\d^2 L_{int}}{\d \Theta(-\textbf{\p})\d \Theta(\textbf{\p})}
\right)\,.
\ee
Note that we chose the current to have support only at low momenta, thus it vanishes when multiplied by the term $d P^\Lambda/d \Lambda$ which has support only near the cutoff $\Lambda$. In the spirit of TSPT we use the effective action ansatz 
\be
\label{eq:anstspt}
 L_{int} = -\sum_{n=1}^\infty\frac{1}{n!} \int_{\q_1}...\int_{\q_n}\Gamma^\Lambda_n(\q_1,...,\q_n)
 \Theta(\q_1)...\Theta(\q_n)\,.
\ee
Plugging this into the renormalization group equation \eqref{eq:rg} yields
\be
\label{eq:rg2}
\begin{split}
 \Lambda\frac{\d \G^\Lambda_n}{\d \Lambda}=&
 \frac{1}{2}\int_\p\L\frac{\d P^\Lambda(\e;p)}{\d \L}\Bigg(
-\G_{n+2}^\Lambda(\q_1,...,\q_n,\p,-\p)\\
&+\sum_{m=1}^{n+1}\sum_{i_1<i_2<...<i_{m-1}}
\G^\Lambda_m(\q_{i_1},...,\q_{i_{m-1}},\p)
\G^\Lambda_{n+2-m}(\q_{i_m},...,\q_{i_{n+m-2}},-\p)
\Bigg)\,.
\end{split}
\ee
These equations cannot be solved exactly as the equation for the $n$-th vertex involves $n+1$'s and $n+2$'s vertices.
One can, however, find solutions to the RG equations perturbatively 
by expanding over the growth factor $g^2$, which corresponds to the TSPT
loop expansion,
\be
\label{eq:L-dep}
\G^\L_n=\sum_{l=0}^\infty g^{2(l-1)} \G^{(l),\L}_n\,,
\ee
where $l$ denotes the loop order. The power spectrum is simply given by $P_L(\e;k)=g^2\bar P(k)$.
Plugging this decomposition into Eq.~\eqref{eq:rg2}, and striping off the delta-functions,
one arrives at the final form for the RG - equations,
\be
\label{eq:rg3_0}
\begin{split}
&\frac{\d \G'^{(l),\L}_n}{\d \L}=
 \frac{1}{2}\int_\p\frac{\d \bar P^\L(p)}{\d \L}\Bigg(
-\G'^{(l-1),\Lambda}_{n+2}(\q_1,...,\q_n,\p,-\p)\Bigg)\\
&+\frac{1}{4}\sum_{m=1}^{n+1}\sum_{i_1<i_2<...<i_{m-1}}\sum_{l'=0}^{l}
\L\frac{\d \bar P^\L(|\sum_{A=1}^{m-1}\q_{i_A}|)}{\d \L}\times \\
&\Big[\G'^{(l'),\L}_m(\q_{i_1},...,\q_{i_{m-1}},-\sum_{A=1}^{m-1}\q_{i_A})
\G'^{(l-l'),\L}_{n+2-m}(\q_{i_m},...,\q_{i_{n+m-2}},-\sum_{A=m}^{n+m-2}\q_{i_A})\\
&+\G'^{(l-l'),\L}_m(\q_{i_1},...,\q_{i_{m-1}},-\sum_{A=1}^{m-1}\q_{i_A})
\G'^{(l'),\L}_{n+2-m}(\q_{i_m},...,\q_{i_{n+m-2}},-\sum_{A=m}^{n+m-2}\q_{i_A})
\Big]
\,.
\end{split}
\ee
Renormalization of the composite operator $\delta[\Theta;\eta]$ is straightforward.
We  just need to add it to the effective action ansatz~\eqref{eq:anstspt}
\be
L_{int}\to L_{int} + \sum_{n=1}^\infty \frac{1}{n!}\int_\k J_\delta(-\k) \int_{\q_1}...\int_{\q_n} K_n^\Lambda(\q_1,...,\q_n;\k)\Theta(\q_1)...\Theta(\q_n)  \,,
\ee
plug it into Eq.~\eqref{eq:rg} and take a derivative w.r.t. $J_\delta$.
This will generate an RG flow for kernels $K_n^\Lambda$
similar to the RG flow of $\Gamma_n$. Just like in the case of $\Gamma_n$,
this RG flow can be solved perturbatively.

Several comments are in order.

1. The equations \eqref{eq:rg3_0} fix the $\Lambda$-dependence of the vertices in the Wilsonian effective action 
so that all physical observables do not depend on the cutoff. This, however, does not specify the vertices completely. For that we need to set up 
initial conditions for the RG equations at any given order in $g^2$.
To that end we need to match the low-energy TSPT correlation functions
to n-point functions of the EFTofLSS. We will discuss this matching in detail shortly.

2. From Eq.~\eqref{eq:L-dep} we observe that all the counterterms are manifestly local in time. This can be contrasted with the usual EFT of LSS 
where locality in time of the final correlation functions 
emerges only \textit{a posteriori}.

\begin{figure}
\includegraphics[width=0.99\textwidth]{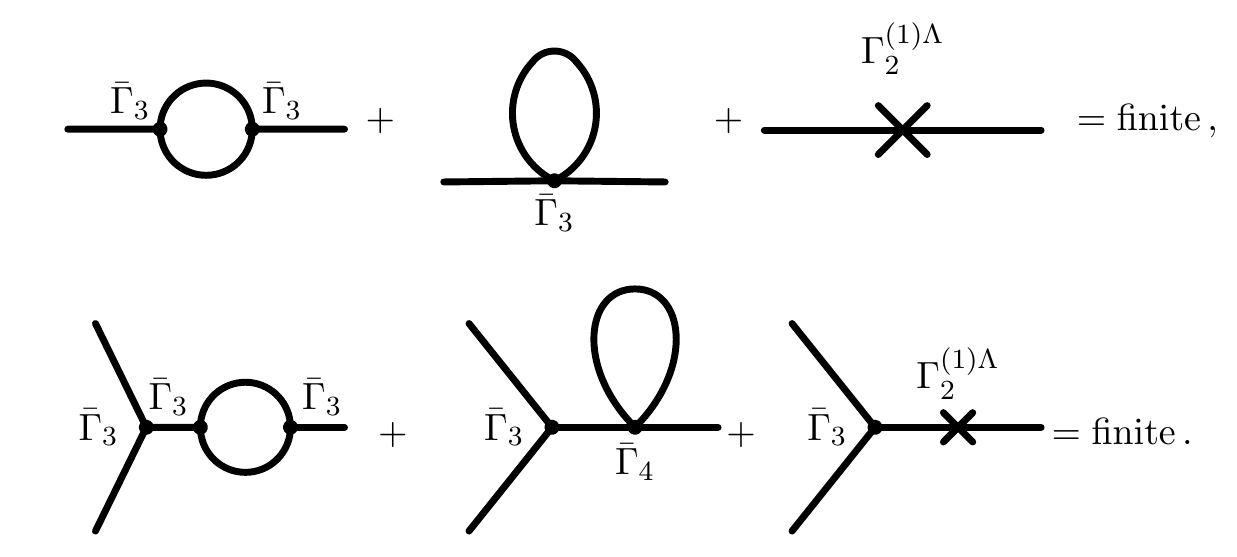}
\centering
\caption{
Diagrammatic representation of the systematic UV renormalization
in TSPT: lower loop order diagrams embedded in higher order ones
(nested divergences) do not require new counterterms.
They are removed by the same counterterms
that renormalize lower order 1PI correlation functions.
}
\label{fig:ctr1}
\end{figure}

3. Renormalization of the connected correlation functions $\G_n$
has an important technical advantage: 
it guarantees that our renormalization procedure in consistent in
the Bogoliubov-Parasiuk-Hepp-Zimmermann sense.
That means the counterterm that cancels a divergence from a certain 1-particle irreducible (1PI) diagram
will also cancel similar divergences that appear in more
complicated diagrams  
embedding this 1PI graph.
Therefore, once a particular 1PI diagram is renormalized, one should not worry about divergences 
which may appear in more complex graphs involving this diagram, --- they 
must be canceled 
by the same 1PI counterterm inserted in an analogous graph. 
This procedure is illustrated in Fig.~\ref{fig:ctr1}.

\begin{figure}
\includegraphics[width=0.99\textwidth]{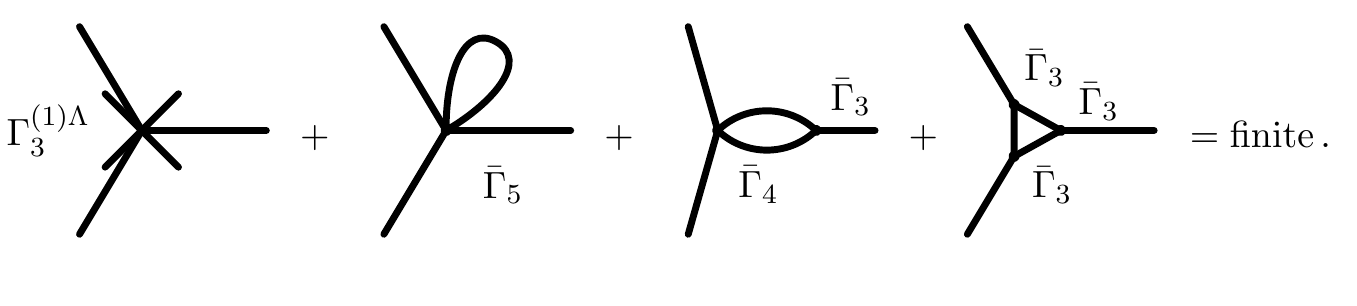}
\centering
\caption{
Diagrammatic representation of the systematic UV renormalization
in TSPT: new counterterms are required only for the 1-particle 
irreducible correlation functions.
}
\label{fig:ctr2}
\end{figure}

As a consequence of the systematic 
renormalization, new counterterms are required 
only for the connected correlation functions, see Fig.~\ref{fig:ctr2}.
Note that this is not the case in the usual Eulerian EFTofLSS~\cite{Abolhasani:2015mra}. 

\subsubsection*{Tree-level RG matching}

It is reasonable to assume that the tree-level 
vertices ($\sim g^{-2}$) for momenta $k<\L$ 
coincide with those derived from PPF.
This is natural as the difference between PPF and real dynamics 
should only appear in loop calculations.
Thus, we can demand that the tree-level vertices reproduce the ones from PPF TSPT 
in the limit $\L \to \infty$. 
An explicit calculation shows that $\Gamma'^{(0),\L}_3(\k_1,\k_2,\k_3) =\bar \Gamma'_3(\k_1,\k_2,\k_3)$ and 
\be 
\label{eq:g40}
\begin{split}
&\Gamma'^{(0),\L}_4(\k_1,\k_2,\k_3,\k_4)=\bar \Gamma'_4(\k_1,\k_2,\k_3,\k_4)\\
&-\left(H(k_{12}-\L)\bar \G'_3(\k_1,\k_2,-\k_{12})\bar P(k_{12})
\bar \Gamma'_3(\k_3,\k_4,-\k_{34})+\text{cyc.}\right)\,.
\end{split}
\ee
Note that the $\Lambda$ dependence appears 
for the first time in the tree-level 
4-point vertex.

\subsubsection*{One-loop RG matching}

The one-loop vertices ($\sim g^{0}$) must cancel the $\L$-dependence
appearing from the loop integrals, hence we call them ``counterterms".
In order to account for physical effects of short modes these counterterms should also have finite, $\L$-independent contributions, specified by the RG 
initial conditions. One way to fix them is to match 
the one-loop renormalized vertices to the TSPT vertices
obtained when the dynamical kernels $I_n$ from PPF 
are replaced by those of the EFTofLSS. 
The first non-trivial counterterm that we find is the one-point
function that cancels the tadpole. This counterterm can be
identified with $C_1$ in the TSPT for the PPF equations.
Solving the RG equations for the the power spectrum, we find the following one-loop
counterterm,
\be
 \begin{split}
 &\G'^{(1),\L}_2(\k,-\k) = \G'^{(1),\text{infinite}}_2(\k,-\k;\Lambda)+\G'^{(1),\text{finite}}_2(\k,-\k)\\
 & =\frac{1}{2}\int_\q\bar P(q) \bar \G'_4(\q,-\q,\k,-\k)
 H(q-\L)\\
 &-\frac{1}{2}\int_\q\bar P(q)\bar P(|\k-\q|) \bar \G'_3(\q,-\q+\k,-\k)
\bar \G'_3(-\q,\q-\k,\k) 
H(q-\L)\\
&+C^{(1)}_2(\k,-\k)+\G'^{(1),\text{finite}}_2(\k,-\k)\,.
 \end{split}
\ee
The first two terms in the r.h.s. above are simply minus the UV limits 
of the one-loop integrals. Their presence guarantees that the UV divergences
of the loop expansion would be canceled for any initial power spectrum.
The third term in the r.h.s. is the TSPT PPF counterterm $C_n$, 
which we included in order to reproduce the PPF result in the $\Lambda\to\infty$
limit.\footnote{Inclusion of this term in $\G'^{(1),\L}_2(\k,-\k)$ is optional. It can be considered as a particular renormalization scheme.}
The last 
term $\G'^{(1),\text{finite}}_2(\k,-\k)$ is the finite counterterm
that captures the physical backreaction of short scale modes.
The simplest way to reproduce the one-loop EFT corrections is to 
use the UV-inspired approach that assumes that the $k$-dependence 
of the finite counterterm matches that of the infinite counterterm part,
\be
\begin{split}
 & \G'^{(1),\text{infinite}}_2(\k,-\k;\Lambda)= 
\mathcal{O}(1)\times\frac{k^2}{P_{L}(k)}\int_{\q,~q\geq \Lambda} \frac{P(q)}{q^2} \\
\Rightarrow \quad & \G'^{(1),\text{finite}}_2(\k,-\k) =\frac{\alpha(\eta)k^2}{P_L(k)k_{\rm NL}^2}\,,
\end{split}
\ee
where the coefficient $\alpha$ needs to be fixed by observations.
Formally, the above equation reproduces the 
power spectrum correction due to the effective stress tensor $\sim k^2P_{11}$.
As mentioned above, however, a more systematic 
procedure would be to match the finite counterterms to the 
EFTofLSS n-point functions. This way it is guaranteed that all physical
EFT corrections are taken into account.

\section{Comparison with Data}\label{sec:data}

As we have discussed earlier, IR resummation 
and UV counterterms are necessary in order to 
describe the actual data. In the context of dark matter 
we show here that the one- and two-loop EFT model 
allows to noticeably increase the range of validity of the 
analytic description and fit the N-body simulation data 
down to scales where naive phenomenological models, like SPT, break down. 

\begin{figure}
\includegraphics[width=0.49\textwidth]{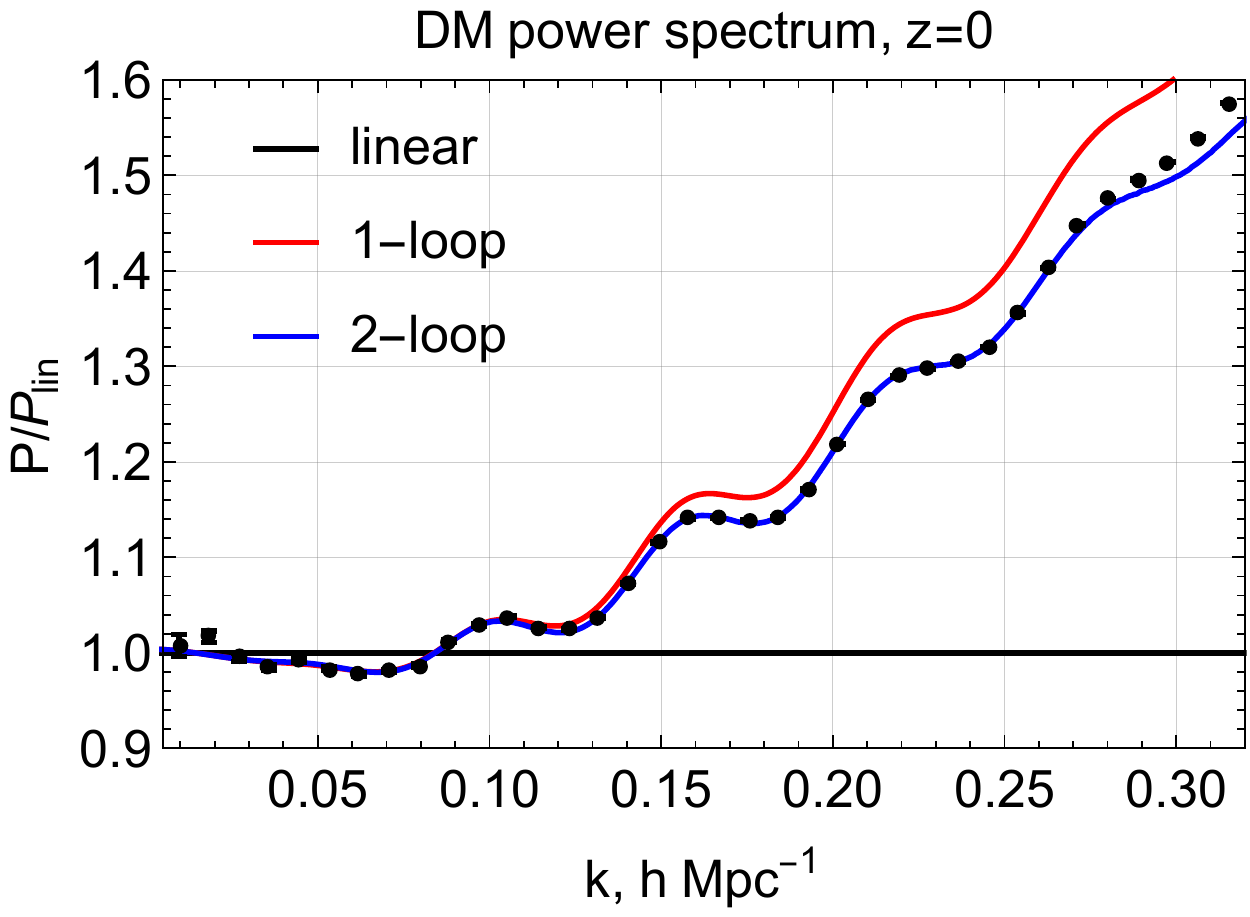}
\includegraphics[width=0.49\textwidth]{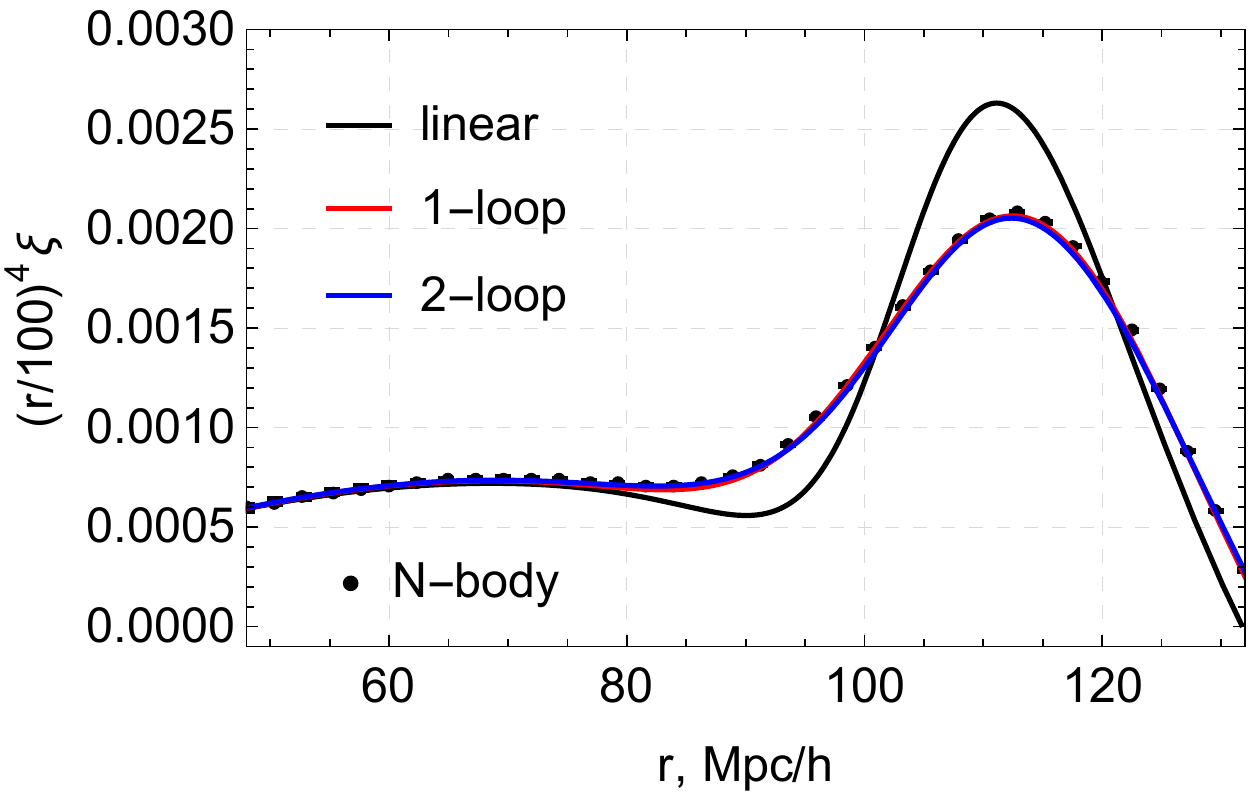}
\centering
\caption{
\textit{Left panel:}
matter power spectra normalized to the linear theory prediction $P_{\rm lin}$. 
We show the fully non-linear spectrum extracted from the
Horizon Run (HR) N-body simulation (dots) at $z=0$, 
the linear theory curve (straight black line),
and the IR-resummed EFT model at one and two loop orders (in red and blue, respectively).
\textit{Right panel:} The position space 2-point correlation
function for the same models against the HR data.
}
\label{fig:fits}
\end{figure}

In Fig.~\ref{fig:fits} we show the results for the matter power spectrum of the Horizon Run 
N-body simulations at $z=0$. We see that linear theory breaks down 
at $k\approx 0.03~\hMpc$, while the one and two loop IR-resummed 
EFT models fit the data up to $k\approx 0.12~\hMpc$ 
and $k\approx 0.27~\hMpc$, respectively. We see that both the overall slope and the 
shape of the BAO wiggles are well captured by the EFT. 
In order to emphasize the success of IR resummation we show in
the right panel of Fig.~\ref{fig:fits} the results 
for the position space two-point correlation function.
We see that the shape of the BAO peak is now accurately captured,
which can be contrasted with the linear theory and PPF approximation results~\ref{fig:bao1}.

\section{Galaxy Bias and Redshift Space Distortions}\label{sec:rsd}

So far our discussion has been limited to the case of dark matter
in physical coordinate space. What we observe in spectroscopic galaxy 
surveys are actually galaxies, which trace matter in a nonlinear fashion. 
The coordinates of these galaxies are reconstructed using 
their observed redshifts,
which are contaminated by peculiar velocities.
Both effects represent extra sources of non-linearity,
which can be systematically incorporate
in the EFTofLSS  theory model.
In this section we will discuss 
these effects very briefly. An interested reader can find 
more details in comprehensive reviews~\cite{Desjacques:2016bnm,Baldauf:2020bsd} and original works~\cite{Senatore:2014eva,Senatore:2014vja,Lewandowski:2014rca,
Assassi:2014fva,
Mirbabayi:2014zca,
Lewandowski:2015ziq}.

\subsection*{Galaxies in the EFT}

The relationship between the galaxy density
and the underlying matter field is called ``galaxy bias.''
In the EFT framework one can describe this relationship
entirely based on the symmetries and the decoupling argument. 
On general grounds, 
the observed galaxy density on large scale can only be a function 
of the long-wavelenght density and the tidal field $t^{ij}\propto \d_i\d_j\Phi-(1/3)\delta_{ij}\Delta \Phi$. Since the time-evolution
of small scales is slow, we need to include memory effects.
To that end one should write down a time integral that involves 
all possible operators consistent with symmetries and
built from 
the long-wavelength degrees of freedom, taken
along the fluid trajectory,
\be
\label{eq:bias}
\begin{split}
&\delta_g = \int^\tau d\tau'\mathcal{H}(\tau')~\Big[K_{\delta}(\tau,\tau')\delta(\x_{\rm fl}(\tau'),\tau')+K_{\delta^2}(\tau,\tau')\delta^2(\x_{\rm fl}(\tau'),\tau')\\
&+K_{t^2}(\tau,\tau')t_{ij}t^{ij}(\x_{\rm fl}(\tau'),\tau) 
+ K_{\nabla^2\delta}(\tau,\tau')R_*^2\nabla^2_{\x_{\rm fl}} \delta(\x_{\rm fl}(\tau'),\tau')+...
\Big]\,,
\end{split} 
\ee
where $K_i$ are some unknown evolution kernels and $R_*$ is a typical
length scale associated with galaxy formation physics, $R_*\sim R_{\rm vir}$ on general grounds. The bias expansion must also include
stochastic contributions that are uncorrelated with the long-wavelenght 
matter fields. In perturbation theory the expression~\eqref{eq:bias}
can be simplified in the same way as we treated the time non-locality
of the effective stress tensor. In particular, at leading order 
one would just get 
\be
\label{eq:bias2}
\begin{split}
\delta_g & = 
\int^\tau d\tau'~K_{\delta}(\tau,\tau')\delta(\x_{\rm fl}(\tau'),\tau')\\
&=
\left[\int^\tau d\tau'~K_{\delta}(\tau,\tau')\right]\delta(\x,\tau)
+\left[\int^\tau d\tau'~K_{\delta}(\tau,\tau')(\tau'-\tau)\right]\frac{D}{D\tau}\delta(\x,\tau)+...\\
&=b_1 \delta^{(1)}(\x,\tau)\,,
\end{split} 
\ee
which is called the linear bias relation. The linear bias 
parameter $b_1$ is a Wilson coefficient, which has to be retrieved from data.
Note that in general the bias coefficients receive UV contributions from 
loops and need to be renormalized.
The EFTofLSS allows one to go beyond the leading order bias expansion~\eqref{eq:bias2} in a systematic and consistent fashion.
A similar progress has been made for baryonic 
effects~\cite{Lewandowski:2014rca,Braganca:2020nhv}.

\subsection*{Redshift space distortions}

The observed coordinates of a galaxy are two angles fixing the position on the sky and a redshift.
If the galaxies were moving exactly with the Hubble flow,
their redshift could be unambiguously converted into a radial distance.
In reality, however, each galaxy has a peculiar velocity w.r.t.
the Hubble flow, i.e. its apparent position is a distorted
version of the actual position. This effect is called redshift-space distortions. 

Neglecting relativistic effects, and assuming
that the galaxy is far away from us, 
the distorted position $\s$ (called a position 
in ``redshift space'') is given by the actual position $\x$ in configuration
space plus a velocity-dependent correction,
\be
x^i_{\rm RS} = x^i +\frac{v^j \hat z_j}{\HH} \hat z_i\,,
\ee
where $\hat z_i$ is the line-of-sight direction unit vector which 
we choose to 
coincide with the z-axis.
Since a galaxy distribution is simply  
reshuffled, the total mass remains intact, which allows us to  
take advantage of mass conservation,
\be
(1+\delta^{\text{(RS)}}_g(\x_{\rm RS}))d^3x_{\rm RS}= (1+\delta_g(\x)) d^3x\,.
\ee
Using the Jacobian of the map $\x\to \x_{\rm RS}$, the galaxy density 
in redshift space can be simply expressed as 
\be
\label{eq:rsdmapp}
 \delta^{\text{(RS)}}_g(\k)=\delta_g(\k) + \int d^3x~e^{i\k\x}\left[
\exp\{i k^i \hat z_i (v^j \hat z_j)/\HH\}-1
 \right][1+\delta_g(\x)]\,.
\ee
Expanding the r.h.s. to linear order and using 
the linear theory results $v^i=-f\HH [\d_i/\Delta]\delta$, $\delta_g = b_1\delta$,
 we get the famous Kaiser formula~\cite{Kaiser:1987qv},
\be 
\delta^{\text{(RS)}}_g(\k) = \left(b_1 + f\frac{k_z^2}{k^2}\right)\delta(\k)\,.
\ee
In order to include non-linear corrections 
we need to retain
higher powers of the velocity field in Eq.~\eqref{eq:rsdmapp}.
This will generate contact operators that have uncontrolled UV 
sensitivity. This can be removed by coarse-graining the 
perturtbative Taylor expansion of~\eqref{eq:rsdmapp},
which will produce new effective operators due to smoothing 
just like the smoothing of the effective dark matter stress-tensor 
produced the effective sound speed and viscosity. This procedure 
can be systematically carried out to any desired perturbative order
in the EFT.

\section{Summary and Outlook}\label{sec:sum}

In this Chapter we have presented the basics of the 
EFTofLSS. We have shown that the description of the 
observed 
galaxy distribution 
consists of three main ingredients: non-linearities 
in the underlying dark matter fluid, 
non-linearities in the bias expansion, 
and non-linearities in the redshift space distortions.

In this Chapter, we have mostly focused on the first 
ingredient: the non-linear evolution of dark matter 
in the perturbative regime. We have shown that the correct and 
self-consistent dark matter model in this case 
is a non-ideal fluid with some effective stress-tensor. 
This stress tensor captures the long-range effects of 
short scale (UV) degrees of freedom. We have also 
discussed in detail the non-linear evolution of 
the BAO, which requires a non-perturbative treatment.
This treatment can be developed in the context of
TSPT. In addition, TSPT provides certain 
insights into the UV renormalization procedure.

Once all necessary
IR and UV effects are properly accounted for,
we find perfect agreement between N-body data and 
the EFTofLSS model for the matter power spectrum.
This is not a coincidence, as the EFTofLSS is the
only mathematically consistent way to build 
a non-linear cosmological perturbation theory
for matter and galaxies.
A key advantage of this approach is its systematicity,
i.e. the ability to account for \textit{all}
non-linear clustering effects in a controlled fashion.
Thus, the EFTofLSS is a powerful and efficient theoretical tool
for precision large-scale structure calculations.

We have touched upon bias and RSD, but very briefly. 
We refer the interested reader to reviews for more detail. 
There are many other topic that we have not mentioned
due to the brevity of this Chapter. This include 
Lagrangian EFTofLSS~\cite{Porto:2013qua,Vlah:2015zda,Vlah:2015sea,Chen:2020fxs,Chen:2020zjt,McEwen:2016fjn},
efficient 
evaluation of loop integrals for the EFTofLSS~\cite{Simonovic:2017mhp,Chudaykin:2020ghx}, 
field level EFT~\cite{Schmidt:2018bkr}, 
higher order statistics~\cite{Philcox:2021hbm,Ivanov:2021kcd,Philcox:2021kcw,Philcox:2022hkh}, 
primordial non-Gaussianity~\cite{Assassi:2015fma,Assassi:2015jqa,Cabass:2022wjy,Cabass:2022ymb,Vasudevan:2019ewf},
massive neutrinos~\cite{Blas:2014hya,Senatore:2017hyk,Chudaykin:2019ock}, 
as well as applications of the EFTofLSS to
the actual galaxy clustering data (e.g.~\cite{Ivanov:2019pdj,DAmico:2019fhj,Ivanov:2019hqk,Philcox:2020vvt,Ivanov:2020ril,Chudaykin:2020hbf,Chen:2021wdi,Ivanov:2021zmi,Ivanov:2021fbu}). 
We hope to explore these 
topics elsewhere in future.

\end{document}